\documentclass[12pt]{article}
\usepackage{amssymb,epsf}
\def\lsim{\mathrel{\rlap {\raise.5ex\hbox{$ < $}}
{\lower.5ex\hbox{$\sim$}}}}
\def\gsim{\mathrel{\rlap {\raise.5ex\hbox{$ > $}}
{\lower.5ex\hbox{$\sim$}}}} 
\def\sqr#1#2{{\vcenter{\vbox{\hrule height.#2pt

        \hbox{\vrule width.#2pt height#1pt \kern#1pt

           \vrule width.#2pt}

        \hrule height.#2pt}}}}

\def\lsim{{\displaystyle
{{\raise-8pt\hbox{$ <$}}
\atop{\raise5pt\hbox{$\sim$}}}}}
\def\gsim{{\displaystyle
{{\raise-8pt\hbox{$ >$}}
\atop{\raise5pt\hbox{$\sim$}}}}}
\def\slsim{{\displaystyle
{{\raise-8pt\hbox{$\scriptstyle <$}}
\atop{\raise5pt\hbox{$\scriptstyle \sim$}}}}}
\def\sgsim{{\displaystyle
{{\raise-8pt\hbox{$\scriptstyle  >$}}
\atop{\raise5pt\hbox{$\scriptstyle \sim$}}}}}
\newskip\humongous \humongous=0pt plus 1000pt minus 1000pt

\newcommand{\sumpf}[0]{\sum_{(H^{\rm f},G^{\rm f})}\! \! \! \!
{\raise
4pt
\hbox{$'$}}\,}

\newcommand{\sump}[0]{\sum_{(H,G)}\! \! {\raise 4pt \hbox{$'$}}\,}

\def\bs{\begin{subequations}}
\def\es{\end{subequations}}
\catcode`\@=11
\newcount\hour
\newcount\minute
\newtoks\amorpm
\hour=\time\divide\hour by60
\minute=\time{\multiply\hour by60 \global\advance\minute by-\hour}
\edef\standardtime{{\ifnum\hour<12 \global\amorpm={am}%
        \else\global\amorpm={pm}\advance\hour by-12 \fi
        \ifnum\hour=0 \hour=12 \fi
        \number\hour:\ifnum\minute<10 0\fi\number\minute\the\amorpm}}
\edef\militarytime{\number\hour:\ifnum\minute<10 0\fi\number\minute}
\def\draftlabel#1{{\@bsphack\if@filesw {\let\thepage\relax
   \xdef\@gtempa{\write\@auxout{\string
      \newlabel{#1}{{\@currentlabel}{\thepage}}}}}\@gtempa
   \if@nobreak \ifvmode\nobreak\fi\fi\fi\@esphack}
        \gdef\@eqnlabel{#1}}
\def\@eqnlabel{}
\def\@vacuum{}
\def\draftmarginnote#1{\marginpar{\raggedright\scriptsize\tt#1}}
\def\draft{\oddsidemargin -.2truein
        \def\@oddfoot{\sl preliminary draft \hfil
        \rm\thepage\hfil\sl\today\quad\militarytime}
        \let\@evenfoot\@oddfoot \overfullrule 3pt
        \let\label=\draftlabel
        \let\marginnote=\draftmarginnote
   \def\@eqnnum{(\theequation)\rlap{\kern\marginparsep\tt\@eqnlabel}%
\global\let\@eqnlabel\@vacuum}  }
\def\subequations{\refstepcounter{equation}%
  \edef\@savedequation{\the\c@equation}%
  \@stequation=\expandafter{\theequation}
  \edef\@savedtheequation{\the\@stequation}
  \edef\oldtheequation{\theequation}%
  \setcounter{equation}{0}%
  \def\theequation{\oldtheequation\alph{equation}}}

\def\endsubequations{\setcounter{equation}{\@savedequation}%
  \@stequation=\expandafter{\@savedtheequation}%
  \edef\theequation{\the\@stequation}\global\@ignoretrue
  \vspace*{-12pt} \\}
\def\bs{\begin{subequations}}
\def\es{\end{subequations}}
\relax
\def\Im{\,{\rm Im}\, }

\def\thefootnote{\fnsymbol{footnote}}
\def\be{\begin{equation}}
\def\ee{\end{equation}}
\def\ba{\begin{eqnarray}}
\def\ea{\end{eqnarray}}

\def\ee{\end{equation}}
\def\bea{\begin{eqnarray}}
\def\eea{\end{eqnarray}}

\newcommand{\uarrw}[0]{\mathrel{
{\raise.5ex\vbox{\hrule width 1cm}\hskip-6pt\rightarrow}}}
\def\thebibliography#1{%
\vskip 0.5cm \centerline{\bf References}
\list{%
[\arabic{enumi}]}{\settowidth\labelwidth{[#1]}
\leftmargin\labelwidth
\advance\leftmargin\labelsep
\usecounter{enumi}}
\def\newblock{\hskip .11em plus .33em minus .07em}
\sloppy\clubpenalty4000\widowpenalty4000
\sfcode`\.=1000\relax}

\renewcommand{\theequation}{\arabic{section}.\arabic{equation}}

\renewcommand{\section}{\setcounter{equation}{0}\@startsection%
{section}{1}{0mm}{-\baselineskip}{0.5\baselineskip}%
{\normalfont\normalsize\bfseries}}

\renewcommand{\subsection}{\@startsection%
{subsection}{2}{0mm}{-\baselineskip}{0.5\baselineskip}%
{\normalfont\normalsize\slshape}}

\renewcommand{\subsubsection}{\@startsection%
{subsubsection}{2}{0mm}{-\baselineskip}{0.5\baselineskip}%
{\normalfont\normalsize\slshape}}

\begin{document}
\renewcommand{\theequation}{\arabic{section}.\arabic{equation}}
\begin{titlepage}
\begin{flushright}
HU-EP-02/31 \\
hep-th/0207195 
\end{flushright}
\begin{centering}
\vspace{1.0in}
\boldmath
{ 
\bf \large Entropy, String Theory, and our World$^\dagger$
}
\\
\unboldmath
\vspace{1.7 cm}
{\bf Andrea Gregori}$^1$ \\
\medskip
\vspace{.4in}
{\it  Humboldt-Universit\"at, Institut f\"ur Physik}\\
{\it D-10115 Berlin, Germany}\\

\vspace{3.2cm}
{\bf Abstract}\\
\vspace{.2in}
\end{centering}
We investigate the consequences of two assumptions
for String ( or M) Theory, namely that: 1) all coordinates
are compact and bound by the horizon of observation,
2) the ``dynamics'' of compactification is determined by the ``second
law of thermodynamics'', i.e. the principle of entropy.  
We discuss how this leads to a phenomenologically consistent
scenario for our world, both at the elementary particle's and at
the cosmological level, without any fine tuning or further ``ad hoc''
constraint.
\vspace{4cm}

\hrule width 6.7cm
\noindent
$^\dagger$Research supported by the EEC under the contract 
EU HPMF-CT-1999-00396.\\
$^1$e-mail: agregori@physik.hu-berlin.de

\end{titlepage}
\newpage
\setcounter{footnote}{0}
\renewcommand{\thefootnote}{\arabic{footnote}}

\tableofcontents

\vspace{1.5cm} 

\noindent

\section{Introduction}
\label{intro}

Since the time of its introduction, and subsequent interpretation as  
a viable quantum theory of gravity and of the other interactions,
there have been innumerable attempts to make of string theory
a ``true'' theory, and not just a complicate machinery able to
produce whatever kind of spectra through a plethora of allowed
compactifications (although never precisely ``the'' spectrum 
one would like to find). It seems that one of the main problems
is the one of providing this theory with a dynamics. The classical way
this problem is phrased and solved is through the introduction
of a ``potential'' in the effective action of the theory.
Indeed, as it is, String Theory looks like an ``empty'' theory,
comparable to what could be in quantum mechanics
having the kinetic term of the harmonic oscillator, and therefore been
able to solve for the ``asymptotic states'', without knowing
what are the interactions of these states.
In the case of string theory, it is already quite hard to find
an appropriate formulation of this problem. The reason is that
the usual Lagrangian approach treats space and time as ``external
parameters'', while in String Theory these are fields themselves.
And it is correct that things are like that: as a quantum theory of gravity
String Theory must contain also a solution for the cosmological
evolution of the Universe. In this perspective, one may worry
about whether it does make any sense at all to speak in terms of 
a potential to minimize, in order to ``stabilize'' the moduli
to the solution corresponding to our present world:
what kind of ``stabilization'' has to be looked for in a theory
supposed to describe the evolution of ``everything'', namely
of the space-time itself? And moreover, what do we have to intend
for the ``space-time''? Normally, we are used to consider this as
the natural framework, the ``space of the phases'' in which
to frame our physical problem. As such, the space-time is 
necessarily infinitely extended, because it is by definition the
space in which the theory is allowed to move. But for string theory
space and time are not the frame, they are a set of fields. Where do they move
these fields? What is their ``phase space'', what is the ``parameter''
that enables to label their configurations? 
A first, key observation, is that, if we interpret the string coordinates as
the coordinates of our Universe, then, despite the common ``lore'',
there is no evidence that (at least) four of them are infinitely
extended. All what we know is what, today, is contained inside
our horizon of observation, namely, the portion of Universe
included in a space-time region bounded by its Age.
Why should a theory of our Universe include in its description 
also regions ``causally disconnected'' from our one? 
In these regions there can be anything, 
and whatever there is there, our physics will not be affected.
Assuming that four string coordinates are infinitely extended
is indeed redundant, it is requiring more than what our experimental
observation honestly allows us to say.
Therefore, in this work we will be ``conservative'', and take
the more ``cautious'' approach of considering them compact, and 
limited precisely by the horizon of observation. As we will see,
this apparently ``innocuous'' assumption has indeed deep consequences
on the way we phrase our physical problems. 
From this perspective, the evolution of String Theory is the history
of our Universe, intended as the space-time and the matter and fields 
inside it. The fact that, as it is, String Theory allows for an extremely huge
number of configurations, a priori all equally viable and valid,
suggests us that if there has to be a
``frame'' in which to ``order'' this history, this 
has to be provided by statistics. The ``dynamic principle'' 
that seems to us the more logical and natural is the second law of 
thermodynamics. 
This provides in fact an ordering in physical phenomena, which
can be associated to time evolution, but does not involve directly
a ``classical'' treatment of time. It looks therefore suitable
for a theory in which time is a field. The fundamental notion
on which to base the ordering is therefore ``Entropy''. But here 
the question: what do we have to intend for it? Usually, this
quantity is defined in relation to the amount of disorder, or
the number of ways it is possible to give rise to a certain configuration,
for a system consisting of many degrees of freedom, normally many particles. 
In any case, a ``macroscopic'' definition. In this work we 
will try to extend, or ``adapt'', the definition, in order to include
a treatment also of ``microscopic'' systems, such as 
the configurations of compact string coordinates.
Inspired by the fact that particles can be viewed in some way as
sources of singularities of space-time, i.e. of large scale coordinates,
we will associate to entropy also the treatment of singularities 
of coordinates at a sub-Planckian scale \footnote{In this work
we will identify the (duality-invariant) string scale
with the Planck scale.}. The inclusion of both scales 
is a natural requirement in order to keep
consistency with a fundamental notion of String Theory: T-duality.

In this work we discuss how with this ``dynamical input''
String Theory can be ``solved'' to give a ``history'' of our Universe,
ordered according to the entropy of the various configurations.
In particular, we will see how an expanding Universe is
automatically implied. Starting at the ``origin''
from a ``flat'' configuration, absolutely democratic in all string 
coordinates, the system
attains the minimum of entropy at the Planck scale. At this scale, 
the string space is the most singular one. In this configuration,
T-duality is broken, and many coordinates are ``twisted''.
Precisely the breaking of T-duality allows to chose an ``arrow'' in the
time evolution, and speak of our ``era'' as the one in which 
statistical phenomena act at a macroscopic level. i.e. at scales 
larger than the Planck length.
Our space-time arises as the only part of the string space whose
coordinates are not twisted, and therefore remain free to expand.
The spectrum of what we call elementary particles and fields
is uniquely determined at the time of minimal entropy,
and coincides with the known spectrum of quarks and leptons, and
their interactions. However, in this set up masses don't arise
as a consequence of the interaction with a Higgs field.
Rather, they are a quantum effect related to finiteness of
our horizon of observation, in turn related to the
``Age of the Universe''. The relation between masses and 
this quantity is not simply the one expected by ordinary
quantum field theory of a system in a compact space:
the space-time of string theory is in fact somehow an ``orbifold''
of what we call space-time. Different types of particles and fields
originate from different ``orbifold sectors'', and they 
``feel'' the space-time each one in a different way.  
In a way reminiscent of the difference of perception of space and rotations
between Bosons and Fermions, the space of string theory, made up of
an extended and an internal one,
provides additional degrees of freedom leading to a further
discrimination of matter through a varied spectrum of masses.    
A consequence of these relations is the decay of particle's masses 
with time. Although at the present day this effect is too slow to be
directly detected, it is by itself able to explain the accelerated
expansion of the Universe. Besides the non-existence of the Higgs field,
in fact not needed in order to provide masses in String Theory,
another aspect of the scenario arising in our set up is
the non-existence of supersymmetry, which turns out
to be broken at the Planck scale. Precisely 
a mass gap between the particles and fields we observe and their
supersymmetric partners of the order of the Planck mass
produces the value of the cosmological constant we experimentally observe. 
As we will discuss, this parameter can be interpreted as the manifestation
of the Heisenberg's Uncertainty Principle, at a cosmological scale.
Put the other way round, this Principle could be ``derived''
from the existence of the cosmological constant.
This fact supports the idea that indeed String Theory provides 
a unified description of Gravity and Quantum Mechanics.

Among the strong points of our proposal is 
the natural way several aspects of our world are explained,
and come out necessarily, without any fine tuning: besides
the already mentioned
cosmological constant and accelerated expansion of the
Universe, the correct spectrum of particles and interactions,
with the correct ``prediction'' for the electroweak scale.
In particular, the parity violating nature of the weak
interactions, as well as the breaking of space and time parities,
appear as the effect of a ``shift'' in the space-time coordinates,
necessarily required by the entropy principle. It is noteworthy that,
without this operation on the space-time, the weak interactions
would not be chiral, something we find rather logical: it had to
be expected that the breaking of parity must be due to an operation
on the space-time. 
Another thing that had in some sense to be expected is that
masses, being sources of, and coupled to, gravity, only in the
context of a quantum theory of gravity, as string theory is,
can receive an appropriate treatment and explanation. 
In this set up, they are not ``ad hoc'' parameters, but appear
as a stringy quantum effect.

The paper is organized as follows.
We first introduce and discuss the basic idea, namely the concept
of entropy we consider the most suitable for string theory. We state then
the ``dynamical principles'', according to which we will let the
system to evolve. A proper treatment of these topics
would require a rigorous mathematical framework in which to phrase
the problem of singularities of the string space, something we
don't have at hand. Throughout all the work we will therefore make
a heavy use of the closest thing we know, the language of orbifolds:
this approach is justified by the hypothesis that
this is a good way of approximating the problem, at least in many cases,
that we will discuss. (this however does not mean at all that we believe
the present world to correspond to a string orbifold!).  
Although, owing to the substantial lack of technical tools, our analysis
is in many cases forcedly more qualitative than quantitative,
nevertheless it is still possible to obtain also some quantitative
results, allowing to test, although with a precision limited
to the only order of magnitude, the correctness of certain statements,
and also to make, up to a certain extent, predictions. 
Obtaining the correct order of magnitude with a ``rough'' computation 
and without any fine tuning, for scales which are
$10^{-17}$ (the electroweak scale) or $10^{-61}$ 
(the square root of the cosmological constant) times smaller than 
the Planck scale, having assumed this last as the only fundamental mass scale,
is anyway already quite significant and encouraging.
After having discussed the issues of supersymmetry breaking and
the cosmological constant, we pass in section \ref{SM} to explain how our
``low energy'' world arises. A subsection is devoted to the issue of masses
and their evaluation. In particular, a prediction for the neutrino
masses, necessarily non-vanishing in this set up, is presented.
In the same section we discuss the issue of the fate of the Higgs
field, and the unification of the couplings.
Finally, based on what suggested by the analysis of the previous
sections, in section \ref{comments} we comment about some aspects of
the possible underlying higher dimensional
theory, which could be behind string theory.
In the framework of our proposal, the breaking of supersymmetry
and quarks confinement in our world appear in a natural way,
without having to be ``imposed'' by hand.

Unfortunately, many topics,
and especially this last one, can receive only a qualitative
treatment. The actual world corresponds in fact to a situation 
of string coupling of order one. The analysis we present here
should therefore be taken, from many respects,
more as suggesting a direction of research,
than as really providing the final answer to the problems.
Despite its degree of approximation, we found it however worth
of been presented. We hope to report on more progress in the future.

\section{Expansion/compactification and the second principle of
thermodynamics}
\label{exp}

\subsection{\sl Entropy in String Theory}

The fundamental notion onto which this work is based is Entropy.
We will discuss how this quantity has to be considered
as the basic ingredient  governing the evolution of the Universe, and
at the origin of the existence of our world, as it is.
In order to do that, we need to give an appropriate definition of this 
quantity, suitable for string theory. String theory is in fact a theory 
in which, owing to T-duality, ``microscopic'' and ``macroscopic'' phenomena,
i.e. phenomena which are above/below the string scale, are mixed,
or better, present at the same time.
In a context independent on which
particular string construction we want to consider, i.e. in a
``string-string duality invariant frame'', namely
in a true ``quantum gravity'' frame, the scale discriminating
what are ``micro'' and ``macro'' phenomena is indeed the Planck scale.  
The notion of entropy we have been accustomed to is a ``macro''
aspect, involving statistical fluctuations occurring at a scale 
above the Planck length. Namely, it has to deal with
multi-particle systems. As is known,
particles and fields are sources of space-time singularities.
On the other hand, they originate from ``internal space'' singularities.
The ``internal space'' is a completely string theory notion. 
The distinction between ``internal'' and ``external'' space 
comes out in string theory as a dynamical consequence: the two spaces
start on the other hand ``on the same footing'', and a definition
of entropy suitable for string theory must include in its domain
a treatment of the ``internal'' space as well. Namely, it has to
be able to underly also the theory of
singularities of this part of the string space.   
Essentially, what we need is a definition of entropy compatible
with T-duality. But what is the meaning of ``statistics'' 
for the sub-Planckian scale? It seems to us reasonable to expect
that, as ``macroscopic'' entropy governs singularities in space-time,
by providing a statistical theory of the sources of it, in a similar
way ``microscopic'' entropy should be a statistical theory
of ``compact space'' string singularities \footnote{Although in this work
we consider \emph{all} the string coordinates compact,
we uniform here to the common way of intending the compact space
as the ``non extended one'', namely, the complementary of our space-time.}.  
In this sense, a flat coordinate is more entropic than one 
which has been orbifolded. 
The reason why we expect this is that an orbifold is a space 
in which the curvature is concentrated at certain points. 
From a geometrical point of view, it is therefore
a highly differentiated, singular space. On the other hand,
from a physical point of view
the spectrum of string theory on an orbifold is reacher than on 
a flat space. Therefore, also in this sense
this space corresponds to a more differentiated, less entropic situation.
Along this line, we expect
more entropic configurations to be those which are 
more ``smoothed down''.

\subsection{\sl The ``dynamical principles'' of the theory}

As we discussed in the introduction, String Theory remains
an ``empty'' theory as long as something that we call dynamical principle, 
which should play a role analog to that of the potential
in a Lagrangian theory, is introduced. In the following we will
call our theory either ``String Theory'' , or also ``M-theory'',
depending on whether we want or not to emphasize the fact that also
the coupling of the theory is a coordinate, a priori on the same footing
as the other ones. From any respect, under the two names
we intend anyway the same theory. 
We make the following ``Ans\"{a}tze'' for the evolution of the system.

\begin{itemize}

\item[(1)] As physical observers we have a horizon. Namely, the maximal
distance at which we are able to see corresponds to the
distance covered by light during a time equal to the age of the Universe.
This must also be the maximal distance at which a theory of our "universe"
is able to "see" at the present time.
Moreover, all the ``events'' falling out of our causal region of Universe are
causally disconnected from all what we can see. Outside of our region
there could be \emph{anything}: in any case this would not affect
\emph{our} physics. Therefore, a theory describing \emph{our} Universe
must be self consistent within our causal region, and to not depend
\emph{at all} on, and say nothing about, whatever there could be outside.  
As a consequence, if we impose that String, or better M- Theory, 
is going to describe our physics, by consistency
we must conclude that all its coordinates  
(including the string coupling), are compact, and limited, in units of time,
by the Age of the Universe.

\item[(2)] During its evolution, String Theory is governed by
the ``second law of thermodynamics''. By this we mean that there is a function
${\cal S}$ that monotonically increases/decreases along with time. We
don't know the ultimate reason for its existence. Let's however assume that
such a ``function'' of the string parameters exists. 
This function is ``entropy'', as we defined it in the previous section,
consistently with a theory describing at the same time 
the ``Universe'' both at large and small scale.

\end{itemize}

\noindent
We will now discuss how
the above principles uniquely determine the evolution of the theory toward the
present configuration of space-time, with the spectrum of particles and
fields, and their interactions, as we observe in current experiments.  
A first observation is that, since we are going to consider all
the coordinates compact, at any stage we are allowed to 
talk about (broken or unbroken) T-duality. Indeed, from the fact that any
compactification/decompactification limit of M-theory
corresponds to a string vacuum, we deduce that the theory
is renormalizable. 
Compactness provides in fact a natural cut-off, and since by
T-duality we get anyway a string limit, we learn that
the theory is cut off both in the ultraviolet and in the infrared. 
In other words,  once we understand that talking of eleven dimensions
is only an approximation, equivalent to the removal of the natural
cut-off, we obtain also that we do not have anymore to worry
about ``non-renormalizability'' of this theory. The ``non-renormalizability''
is only an artifact due to our arbitrary removal of its natural cut-off.
We should better speak of ``eleven coordinates'', and talk about eleven
dimensions only for certain practical purposes.
Moreover, since we consider time of the same order of the spatial
coordinates, we are allowed to speak of T-duality also in this coordinate,
intending that an inversion of time corresponds to an inversion
of all length scales \footnote{Instead of time, we could better speak of the
``radius'' of the Universe.}. In this sense, T-duality along the
time coordinate $t$ tells us that ${\cal S}$ has a minimum
at $t=1$ in Planck units. ${\cal S}(t)$ behaves in fact
as ${\cal S}(1 / t)$. Therefore, if for $t > 1$ $\partial_t{\cal S} \geq 0$,
$\partial_t{\cal S} \leq 0$ for $0 < t < 1$.
\begin{figure}
\centerline{
\epsfxsize=12cm
\epsfbox{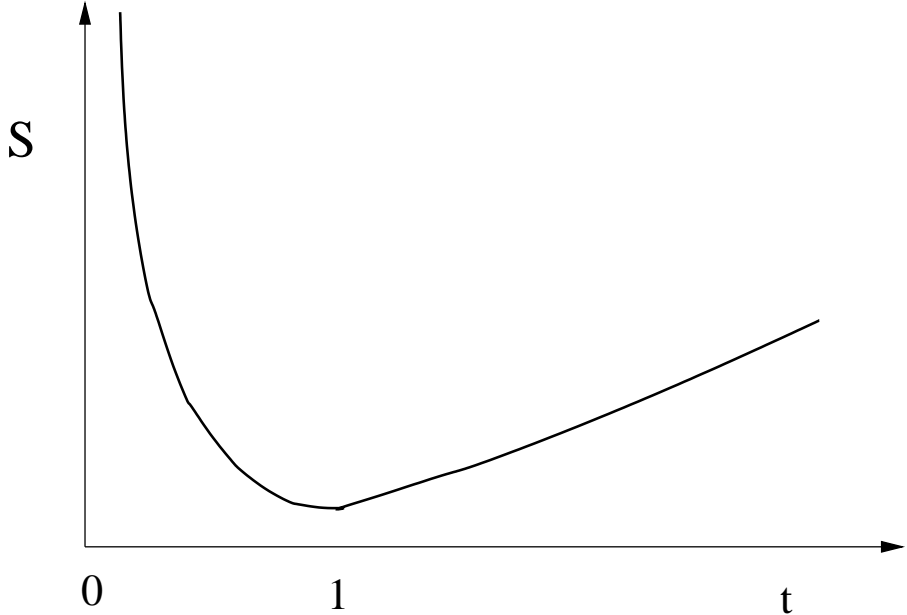}
}
\vspace{0.3cm}
\caption{The approximate behavior of entropy, ${\cal S}$, 
as a function of time, in String Theory.}
\label{Sfig}
\end{figure}
\noindent
Let's make the hypothesis that
at $t=0$, the theory starts in a completely symmetric
configuration, in which statistical fluctuations involve both
the large and the small scale geometry.
As long as T-duality is not broken, ``macro'' and 
``micro'' phenomena, namely phenomena taking place ``above'' or ``below''
the time scale, are equivalent.
As we will see, at the minimum T-duality is broken,
and several string coordinates are twisted.
This means that the theory is not exactly symmetric in $t$ and $1 / t$;
this imposes a choice of ``time direction'', with a consequence also
on the type of statistical phenomena governing the increase/decrease 
of entropy. The breaking
of T-duality has in fact the effect of making inequivalent the two
branches, $t > 1$, and $t^{\prime} \equiv 1 / t > 1$.
For obvious reasons, we make the choice 
that it is  for $t > 1$ that the mass scale of the Universe is below the Planck
scale.  According to this choice, for $t > 1$
the statistical phenomena underlying the modification of entropy
work at a ``macroscopic scale'', namely at the level of observable
particles and extended objects, living in the string coordinates
which are not twisted. Only these, as we will see, are free to expand,
and give, at sufficiently large ages, the impression of infinitely extended
space-time in which we live. If the scale was instead above the Planck scale,
the statistical fluctuations leading to
a change of entropy would act at the string scale, modifying 
also the configuration of the twisted coordinates. This is what happens
for $t < 1$, where, owing to the generically unbroken T-duality,
the statistical fluctuations that lead to
a change of entropy act both above and below the string scale.

From the time of minimal entropy onwards, 
the geometry of the ``internal space''
cannot be ``shaken'' again. Therefore, 
what we observe now are only the ``macroscopic'' statistical
fluctuations, leading to the ordinary definition of entropy.
Notice that, since
for $t < 1$ the ``large'' scale is what for 
$t > 1$ is a small scale, the statistic of different ``large scale
space-time'' configurations for $t < 1$ is what from our point of 
view ($t > 1$) looks as a motion through ``internal space'' configurations.
On the other hand, statistical phenomena underlying the fluctuations 
of the ``small scale'' geometry at $t < 1$
are the primordial of what, in our ``era'',
$t > 1$, are fluctuations of the space-time geometry.
The latter is nothing else than the statistics of a many-particle system, 
because precisely fields and particles are the singularities working 
as sources for the space-time geometry.   
With reference to figure \ref{Sfig}, this means that, by going backwards
in time toward the origin, the internal coordinates are progressively
un-twisted. The theory starts therefore at $t = 0$
from a flat configuration of all the coordinates.

\subsection{\sl The supersymmetry breaking scale}

Let's follow the evolution of ``M-Theory'', starting
from the origin, $t=0$. 
We assume that at the very beginning all the coordinates are
flat. This is the smoothest configuration,
the one that treats all the coordinates in the most democratic way,
and represents therefore the most ``entropic'' situation. We now
allow for some mechanism that generates a ``time'' evolution, through
some fluctuations among the various possible configurations 
\footnote{A first remark is in order:
strictly speaking time is not a parameter, but one of the coordinates
of the theory. What actually determines the ``progress'', or ``time'' 
evolution, is the change of entropy by moving to
different configurations. The fact that this evolution can be ``ordered''
in a sequence of steps happening along the string zeroth coordinate
is made possible by the fact that the latter is free to expand.
This is not at all an obvious fact: for this to happen
it is essential that this coordinate does not get ``twisted'' 
at a certain point.}. We start
therefore flowing toward the right of the X-axis of figure \ref{Sfig}.
For what we said, although we don't know what is the ``real'' size
of the Universe at the beginning, this question has no relevance for
``our'' theory, bounded by our horizon. 
Therefore, not only the ``time'' coordinate but also all the other ones
are at their ``zero''. This means that the theory contains also 
a T-dual part, in which we can consider the space-time infinitely extended.
In particular, it is legitimate to treat the theory by a ``string'' 
approximation, valid whenever the coordinates below the tenth,
seen as coupling from the low dimensional theory, are sufficiently 
small/large. All that to ensure us about the existence of a massless
excitation that we call ``graviton''. This propagates
with the speed of light; 
its existence tells us that, as time goes by, also
our horizon expands at the same speed, in all the directions along
which the graviton is free to propagate. At the very beginning, this means
along all the directions. Statistical fluctuations
progressively introduce singularities leading to more ``curved'' 
configurations. In particular, as we will see, this would lead to
the breaking of T-duality, ultimately allowing us to speak about
``expansion'', as really distinguished from the ``contraction'', and time
evolution in one direction.

In order to analyze the statistical fluctuations and the configurations
they produce, it is convenient to use the language of $Z_2$ orbifolds. 
The reason is that orbifolds 
provide a ``fine'' parametrization of a singular space and
can be constructed also for spaces
whose coordinates are at the Planck length, therefore out of
the range of ``classical'' geometry. Among them, $Z_2$ orbifolds 
are the ``finest'' ones. Only
$Z_2$ projections are in fact allowed to act ``coordinate by coordinate'', and
the maximal amount of singular points in a compact $S^n$ space is
obtained for $S^n \approx \left( S^1 \big/ Z_2 \right)^n$.
Two remarks are however in order: 1) we are not stating here that the
string space \emph{is} an orbifold. Rather, we are here supposing
that $Z_2$ orbifolds are in some sense ``dense'' in the space of
singular configurations, in a way somehow reminiscent of the fact that
the set of rational numbers is dense in the set of real numbers.
In other words, we make the hypothesis that $Z_2$ orbifolds 
constitute a good approximation of the parametrization of 
singular spaces.
2) In analyzing $Z_2$ orbifolds, we will always rely on the 
properties of string orbifolds.

Starting at the ``zero'' with the flat, most entropic situation,
and following the time evolution, for what we said 
we should expect to progress toward less and less entropic configurations,
up to the less entropic situation, realized at time ``1'' in Planck units. 
As we said, this latter can be identified as
the most singular, most ``differentiated'' configuration. 
This means that at that point
the string space is the most singularly curved one.
To be honest, speaking of ``curvature''
is not much appropriate at the Planck scale. The notion of ``space'' 
is in fact  inherited from our classical concept of ``geometry'', 
a notion not quite applicable at the Planck scale.
As we said, this is a further reason why orbifolds are a good
tool for our analysis.
In order to understand how the theory looks like at this point, we
can proceed by ordering
the string configurations according to their ``microscopic'' entropy, 
i.e. from the point of view of a scale ``above'' the Planck mass. 
In terms of $Z_2$ orbifolds, the sequence can be summarized as:
\be
{\cal N}_{11}=1 \; \to \; {\cal N}_{10}=1 \; \to \; {\cal N}_{6} = 1 
\; \to \; {\cal N}_4 = 1 \; \to \; {\cal N}=0 \, .
\label{subr}
\ee
There are indications that
the last step, from ${\cal N}_4 = 1$ to ${\cal N}=0$, is not really an
independent one, the complete breaking of supersymmetry being already
realized at the ${\cal N}_4=1$ level \footnote{A discussion
can be found in Ref~\cite{mayr}.}. 
According to Ref.~\cite{striality}, the reason seems to be that
in ${\cal N}_4=1$ string vacua there is 
always at least one sector that goes to the strong coupling 
\cite{striality} \footnote{This is due to the fact that
in such configurations the so called ``${\cal N}=2$ gauge beta-functions''
are never vanishing. As discussed in Ref.~\cite{striality},
this is a signal of the existence of sectors of the theory
at the strong coupling.}:
this leads to supersymmetry breaking by gaugino condensation.
In section~\ref{SM} we will discuss more in detail the configuration of the
string space at time 1. As we will see, minimization of entropy 
requires the presence of shifts besides the orbifold twists.
For the moment  we are however interested only in the scale of 
supersymmetry breaking, for which it is enough to know just   
the pattern of twists,  Eq. (\ref{subr}).
According to the principles of evolution we have stated, the last step,
the one of maximal twisting, is attained precisely at the Planck scale.
Even though it is possible in principle to imagine that 
one or more coordinates, under casual fluctuations, may get twisted at a
``previous'' scale, namely, before having expanded up to 1,
further fluctuations move the theory away from this ``intermediate''
configurations: such fluctuations act in fact on the ``internal'',
i.e. microscopical, geometry  as long as the scale is above 1.
Any twist above this scale is only temporary, and in the average any
coordinate is free to expand until 1 is reached.
When scale one is reached, the system has attained the maximal allowed
number of coordinate twists,  and from now on
things change dramatically. Owing to the shifts required by minimal entropy,
T-duality is broken. A further expansion brings us into the region
``below'' the Planck scale, not anymore equivalent to the one ``above''.
Below the Planck scale,
statistical fluctuations leading to a change of entropy act on a
``macroscopic'' way. Namely, they are not anymore able to move the
theory away from the singularities of the compact space produced 
up to the scale 1; rather, such fluctuations appear in the way we are used to
see them, involving configurations of the degrees of freedom living
on the untwisted coordinates, the only ones along which interactions
and fields propagate. These are the coordinates that therefore,
being continuously expanding with our horizon, we interpret as 
``space-time''. The other ones, stuck at the Planck length, constitute
the ``internal'' space of the theory.

A cross, indirect check of the correctness of our thermodynamical
considerations is provided by the analysis presented in 
Ref.~\cite{striality}, about the mechanism of non-perturbative
supersymmetry breaking. In Ref. \cite{striality} we argued in fact 
that this phenomenon happens at any time
there is a non-vanishing gauge beta-function, and that, if generically
the coupling of a weakly coupled sector is parametrized by
the moduli $X$ and $Y$, such that ${1 \over g^2} \sim \Im X + \beta \Im Y
{\rm f}(Y)$, for $\Im X$, $\Im Y > 1$
the coupling of the strongly coupled sector seems to be parametrized inversely:
$g^{2 \, \prime} \sim \Im Y {\rm f}(Y) \, + \, \ldots$.
The moduli $X$ and $Y$ are related to the coordinates of the compact space,
and are measured in the duality-invariant
Einstein frame, whose scale is set by the Planck mass.
The fact that the term $\sim \beta \Im Y {\rm f}(Y)$
is at its minimum, $\sim {\rm O}(1)$, when the moduli are at the Planck
scale, indeed suggests that precisely  this is the scale at which it is 
possible to smoothly pass from the weak coupling to
${\cal N}_4=1$/${\cal N}=0$, 
with the consequent splitting into weakly and strongly coupled
sectors. This must therefore be the scale
at which supersymmetry is finally completely broken.
This scale sets also the mass gap between the observed particles
and their supersymmetric partners.
Put in other words, in ``${\cal N}_4=1$'' string vacua 
the masses of the superpartners of the non-perturbatively
broken supersymmetry are tuned by fields, such as the one parameterizing
the string coupling, which are twisted, and ``frozen''at the Planck scale.
At $t=1$  this is the scale of \emph{all} masses, because
all the coordinates of the theory, and therefore all mass parameters,  
are at the Planck scale. 
However, when the system further evolves toward lower scales, namely,
when the horizon further expands, mass terms depending on the ``internal'',
twisted coordinates, remain at the Planck scale, while those depending only
on the coordinates free to further expand with the scale of the Universe,
decrease as the horizon increases. Of this second kind are the masses
of all the particles and fields we experimentally observe: they are
related to the expanding scale, or, in other words , to the Age
of the Universe, through a set of relations that we will discuss
in detail in the following sections. What we can say now is that
these ``low'' masses originate from ``nothing'', i.e. are a 
quantum effect, that from the string point of view is produced by
shifts in the non-twisted, space-time coordinates, as required
by entropy minimization at $t=1$.
These shifts, which also lead to the breaking of C, P, T parities,
lift also the masses of the would-have-been superpartners;
the mass gap $\Delta m$ inside supermultiplets
remains on the other hand the one set at the Planck scale: $\Delta m \sim 
{\cal O}(1)$.

\subsection{\sl The cosmological constant}

We come now to the issue of the cosmological constant.
We have seen that, in the ``string evolution'', the string, or better
M-theory coordinates, have to be considered compact. 
However, we have also seen that, while four
of them, that we consider as the space-time, are free to perpetually
expand, in our branch, $t=1$, the other ones are
``stuck'' at the Planck length. Namely, when measured in Planck units, 
the ``internal''radii are of length ``one''. 
As we have seen, the Planck scale is also the scale of supersymmetry
breaking, and sets the order of magnitude of the mass gap
between particles (fields), and s-particles (s-fields).
Let's see what are the implications of this fact for the cosmological 
constant $\Lambda$. 
In the past, a high mass of the supersymmetric particles was
considered incompatible with the present, extremely small value
($\sim 10^{-122}$ in ${\rm M}^2_{\rm P}$ units) of this parameter
\cite{debern,melch}.
More recently, it has also been proposed that the low value of $\Lambda$
could be justified by the fact that, even in the presence of very heavy
superparticles, supersymmetry could be near the corner in the
gravity sector. In this case the
value of $\Lambda$ would mostly depend on the gravitino mass, that
experimentally cannot be excluded to be very low.
Besides this, a plethora of models able to justify a low value of the 
cosmological constant has been produced. All these models
deal in a way or the other with
effective configurations of membranes and fields. 
We will not review them:
here we just want to show that all these arguments are not  
necessary: the smallness of $\Lambda$ can be seen to be 
\emph{precisely} a consequence of the breaking of supersymmetry 
at the Planck scale.
In order to see what is the present value of $\Lambda$ predicted
in our set up, we can proceed by mapping the configuration
corresponding to the Universe at the present time into a dual
description in terms of weakly coupled string theory. This is possible
because, owing to the fact that several string coordinates
at the present time are very large, we can always
trade one extended space-time dimension for the 
M-theory eleventh coordinate. At least as long as we are interested in just
an estimate of the order of magnitude of $\Lambda$, we can compute the
vacuum energy in the (dual) string configuration at the weak coupling, 
in a lower number of dimensions. $\Lambda$ will then correspond
to a string partition function:
\be
{\cal Z} \approx \int_{\cal F} \sum_i \, (-1)^Q \, 
q^{P^{(i)}_{\rm L}} \bar{q}^{P^{(i)}_{\rm R}} \, ,
\label{zeta}
\ee  
where, as usual, $q = {\rm e}^{2 \, \pi \, {\rm i} \tau}$, ${\cal F}$
indicates the fundamental domain of integration in $\tau$ and $Q$
indicates the supersymmetry charge. 
A mass gap between target-space ``bosons'' and ``fermions''
of order $\Delta P \sim 1$ leads to a value:
\be
{\cal Z} \sim {\cal O}(1) \, .
\label{lambda}
\ee
Although this number is not at all small,
it is precisely what is needed in order to match with
the experimental value of the cosmological constant.
In order to see this, we must consider \emph{what} we have to compare
with what. Namely, consider the usual way the cosmological constant
appears in the effective action:
\be
{\bf S} \approx \int d^4 x \Lambda \, + \, \ldots \; ,
\label{esse}
\ee   
Now, keep in mind that expression (\ref{zeta}), although technically
computed in a weakly coupled theory in less dimensions, it actually
gives the string evaluation of the cosmological constant for the physical
string vacuum. In order to compare it with the effective action,
we must consider that in the expression (\ref{zeta}),
we have integrated over the ghost contribution, or, in other words,
we are in the light-cone gauge, in which only $D-2$ transverse space-time
coordinates appear. The other two can be considered as  
``identified'' with the string word-sheet coordinates.
In order to compare \ref{zeta} with \ref{esse}, we must therefore integrate 
over two space-time coordinates also in \ref{esse}, or in other words
put them at the natural, string-duality-invariant scale, the Planck scale
\footnote{This kind of manipulations make sense because indeed
we consider also the target space coordinates to be compact, so that we can 
``rescale'' them to the Planck length, and consider them of the same order
of the compact, world-sheet coordinates.}.
In order to compare with the string result, expression \ref{esse}
must therefore be corrected to:
\be
{\bf S} \to {\bf \tilde{S}} = \int d^2 x \, V_2 \, \Lambda \, , 
\ee
where
\be
V_2 \, = \, \left( \int^1_0 \int^1_0 d^2 x  \right)  \, \sim \, {\cal O}(1)
\, ,  
\ee   
in Planck units. Finally, we must take into account that the idea of an
infinitely extended space-time is only an approximation, valid in order to
describe the local physics in our neighborhood, but has no
physical justification: as we said, the theory can not in fact know about
what happens in regions causally disconnected from our one, and a 
fundamental theory of our world, namely, of the world we observe, must
be self contained inside our causal region, at our present time.
The value of the string coordinates in the target space
sets our present day horizon. This means that the integration
over the two ``transverse'' coordinates is performed up to the
present-day border of our causal region, namely on a volume determined by the 
distance covered by a light ray since the ``origin'' of the Universe.
Therefore, what has to be compared with ``1'', is: 
\be
\int_{\rm Age~of~Universe} d^2 x \, \Lambda ~~ \approx {\cal T}^2 \, \times \, 
\Lambda \, ,
\ee
where ${\cal T}$ is the Age of Universe. In Planck units, ${\cal T} = {\cal O}
\left( 10^{61} {\rm M}^{-1}_{\rm P} \right)$. Comparison of the string 
prediction,  (\ref{lambda}), with the effective theory parameters in 
(\ref{esse}), requires therefore $\Lambda$ to be of order
$10^{-122} \,  {\rm M}^2_{\rm P}$, as suggested by the most recent 
measurements \cite{debern,melch}.

Notice that this value is exactly the one we would 
obtain by considering that $\Lambda$ is a measure of the ``energy
gap'' of the Universe within the time of its existence, as derived from the
Heisenberg's Uncertainty Principle:
\be
\Delta E \, \Delta t \, \geq \, 1 \, ,
\label{dedt}
\ee
if we use $\Delta t = {\cal T}$. Namely, since the ``time'' length
of our observable Universe is finite, the vacuum energy cannot 
vanish: it must have a gap given by (\ref{dedt}).
$\Lambda$, which has the dimension of an energy squared, is then 
nothing but the square of this quantity\footnote{In other words, 
$\Lambda$ is a pure quantum effect. Since it
is not protected by any symmetry, it is 
naturally generated by ``quantum corrections'', and its size
is set by the natural cut-off of the Universe.}.
We find this very deep: 
the existence of the cosmological constant is nothing but the
manifestation of the Uncertainty Principle on a cosmological scale.
This fact ``reconciles''
Quantum Mechanics with General Relativity.

Notice that, owing to the expansion of the horizon, $\Lambda$ turns out to be
a function of time, and decreases as the Age of the Universe increases.
This is a simple consequence of (\ref{lambda}), or, if one wants, of
(\ref{dedt}).
In particular, this means that the geometry of our space-time
tends to a flat limit. We will come back to this point 
in section~\ref{comments}.

\section{How does the ``Standard Model'' arise}
\label{SM}

Let's consider now what is the low energy world arising in this set up.
As we mentioned, the ``final'' configuration of string theory, 
namely, down on from the Planck scale, is such that there are no
massless matter states. What distinguishes the light, observable particles,
from other string massive excitations that we are not anymore allowed
to interpret as particles, is that the masses of the first ones originate
from the compactness of the space-time, while the others are massive also
in ``true'', infinitely extended, four dimensions, their mass being
related to the internal, twisted coordinates. Therefore, while the
first ones have masses related to the space-time scale, the others 
are ``stuck'' at the Planck scale.

We start here the analysis of the configuration at $t=1$ by
artificially considering the non-twisted coordinates as infinitely
extended. The condition of their compactness will be included
later.  In the spirit of the usual reference to $Z_2$
orbifolds as to the most convenient situations, 
what we have to consider are indeed
the non-freely acting ones: these are in fact those that lead to the
most singular ($\equiv$ less entropic) configuration of the 
``internal'' string space. 
Orbifolds in which matter and field masses are \emph{all} lifted by
shifts in the internal space, namely  the freely acting ones,
represent less singular $\equiv$ more entropic configurations, because
the configuration of the internal space is more ``simple''.
For instance, they don't leave room for a chain of several 
subsequent rank reductions. They are therefore not favored by 
the minimal entropy ($\equiv$ maximal singularity) requirement.

\noindent
\subsection{\sl The origin of three generations}

The less entropic configuration among non-freely acting, $d=4$ 
$Z_2$ orbifolds, is achieved with the so called ``semi-freely acting'' ones.
These orbifolds are those in which, 
although the projections reducing supersymmetry
don't act freely, the number of fixed points is the minimal one.
This condition is achieved by further acting on a non-freely acting orbifold 
with rank-reducing projections of the type discussed in \cite{dvv,6auth,gkr,
striality}. 
They represent therefore the most singular configuration, the one
in which acts the highest number of projections, inside this
kind of spaces. The presence of massless matter is in this case still
such that the gauge beta functions are positive, ensuring thereby
``stability'' of the vacuum \footnote{Stability means here that,
once we focus on this vacuum, the asymptotic states are the good
ones. Namely, the vertex operators indeed describe infrared-free 
objects.}. Let's see in practice what are the ``ideal''
steps of increasing singularity the system encounters when attaining its
most singular configuration.
This description is necessarily only qualitative. Indeed, 
when all the M-theory coordinates are at the Planck scale,
the string coupling is close to one, and to be rigorous we could not think
in terms of perturbative string realizations, such as the orbifolds are.
However, as long as we are interested more in qualitative than quantitative
aspects, we can suppose the eleventh coordinate to be
not exactly at the Planck scale, and then extrapolate the ``result'' to that
scale. Alternatively, it is possible to analyze the
structure of singularities by mapping to a dual, weakly coupled
description, in which the eleventh coordinate is traded by one
space-time coordinate. Although the structure of the spectrum looks
there quite different, the two descriptions are equivalent for
what matters the counting of singularities, and the counting 
of orbifold projections can be done in a rigorous way~\footnote{To be
honest, the physical situation at $t=1$ corresponds to the
``critical point'' in which all the coordinates are of length 1.
In this case, there is no perturbative approach. In order to get informations
about this point, we have to rely on hypothesis of ``continuity'' of the 
solution, that we have under control as soon as at least one coordinate
is larger or smaller than one.}. 
Let's therefore speak in terms of orbifolds. After all, this is not
so different from what one does when analyzes the strong interactions
in terms of free quarks, whose interaction has a coupling with negative
beta-function. 
Starting from the M-theory situation with 32 supercharges,
we come, through orbifold projections, to a string-like situations 
with 16 supercharges and a gauge group of rank 16.
Further orbifolding leads then to 8 supercharges and introduces 
for the first time non-trivial matter states (hypermultiplets).
As we have seen in \cite{striality} through an analysis of
all the three dual string realizations of this vacuum
(type II, type I and heterotic), this orbifold possesses
three gauge sectors with maximal gauge group of rank 16 in each.
The matter states of interest for us are hypermultiplets
in bi-fundamental representations: these are in fact those that
at the end  will describe leptons and quarks (all others are finally 
projected out). As discussed in \cite{striality},
in the simplest case the theory has 256 such degrees of freedom.
The most singular situation is however the one in which,
owing to the action of further $Z_2$ shifts, the 
rank is reduced to 4 in each of the three sectors.
In this case, the number of bi-charged matter states is also
reduced to $4 \times 4 = 16$ \footnote{Notice that these states
never appear as explicitly ``tri-charged''.}. 
These states are indeed the twisted states associated
to the fixed points of the projection that reduces the amount of
supersymmetry from 16 to 8 supercharges.

The next step in the evolution of string theory is a further reduction
to four supercharges (corresponding to ${\cal N}_4=1$ supersymmetry).
If we think in terms of $Z_2$ projections, it is easy to recognize that
in this way we generate a configuration in which
the previous situation is replicated three times. This cannot be explicitly
represented in terms of $Z_2$ orbifolds of the type II string. The best
points of view from which to look at this configuration are those
of the heterotic and the type I string. There, one of the projections
that appeared explicitly on the type II string, the ``Ho\v{r}ava--Witten''
projection, is ``hidden'', i.e. it is fully non-perturbative.  
On the other hand, the coordinates of the theory that explicitly appear 
allow us to see the effects of the further projection. This latter generates, 
together with the previous one, three twisted sectors of the same kind of the
one just seen above. In each of these sectors, there is a similar structure
of the matter states, that now will however be 
multiplets of the ${\cal N}_4=1$ supersymmetry. Therefore, at this stage,
the structure of ``bi-charged matter'' states is triplicated.    
What is interesting is that, in this case, the so-called ``${\cal N}=2$
gauge beta-functions'' are unavoidably non-vanishing. 
According to the analysis of 
Ref. \cite{striality}, this means that there are hidden sectors at the
strong coupling, and therefore that supersymmetry is actually broken
due to gaugino condensation. The supersymmetry breaking appears therefore 
to be non-perturbative, parametrized by the string coupling, or, in general,
by the ``extra'' coordinates of ``M-theory''.

Let's summarize the situation. The initial M-theory underwent 
three projections and now is essentially the following orbifold:
\be
Z_2^{(\rm H.W.)} \times Z_2^{({\cal N}_6=1)} \times Z_2^{({\cal N}_4=1)} \, . 
\label{zzz}
\ee
The labels are intended to remind that the first projection
is essentially the one of Ho\v{r}ava and Witten \cite{hw1}.
For convenience, we distinguished also the other two, 
although all these projections are absolutely equivalent, and
interchangeable. 
The ``twisted sector'' of the first projection gives rise to 
a non-trivial, rank 16 gauge group; the twisted sector of the second
leads to the ``creation'' of one matter family, while after the
third projection we have a replication by 3 of this family,
from the $Z_2^{({\cal N}_4=1)}$ and $Z_2^{({\cal N}_6=1)} \times
Z_2^{({\cal N}_4=1)}$ twisted sectors.
Notice that, as it can be seen on the type II side, where
only the pairs $Z_2^{(\rm H.W.)}$ and $Z_2^{({\cal N}_6=1)}$, or
$Z_2^{(\rm H.W.)}$ and $Z_2^{({\cal N}_4=1)}$ can be explicitly realized,
the products $Z_2^{(\rm H.W.)} \times Z_2^{({\cal N}_6=1)}$
and $Z_2^{(\rm H.W.)} \times Z_2^{({\cal N}_4=1)}$ don't lead to new matter 
families, but rather to doubling of gauge sectors under which the matter 
states are charged; in other words, the product of these operations
leads precisely to the spreading into sectors that at the end of the day
separate into weakly and strongly coupled.
Precisely this fact will allow to interpret the matter states as quarks
\footnote{As we will discuss, the leptons show up as singlets inside 
quark multiplets.}. The product~(\ref{zzz}) 
represents the maximal number of independent twists the 
theory can accommodate. Further projections are allowed, but no further
twists of coordinates. Precisely these twists allow us to distinguish
``space-time'' and ``internal'' coordinates. While the first ones 
(the non-twisted) are free to expand, 
the twisted ones are stuck at the Planck length.
The reason is that the graviton, and as we will see the photon,
in a theory with three twists, live in the (four) non-twisted coordinates.
Precisely the fact that light propagates along these allows us to perceive
them as extended, namely, as our ``space-time''.
Indeed, the physical horizon starts expanding 
as soon as the graviton is created.
This means, from the very beginning. However, before the
point of minimal entropy, the theory possesses ``T-duality'' 
\footnote{this appears for instance through T- and S-dualities of sufficiently
``regular'' heterotic string vacua.}:
expansion and contraction are therefore essentially undistinguished.
When the string space is sufficiently curved, 
T-duality is broken, because the configuration of minimal entropy 
involves also shifts along space-time coordinates . This is what happens at
$t=1$, when, as we will see, also the photon is created.   
Since the process of entropy minimization before
the Planck scale leads to the
maximal twist (\ref{zzz}), the further expansion of the 
horizon at length scales larger than the Planck length, 
namely, in our ``era'', takes place
only along four coordinates, those in which matter and its interactions
live.

Let's now count the number of degrees of freedom of the matter states.
We have three families, that for the moment are absolutely identical:
each one contains $16 = 4 \times 4$ chiral fermions (plus their
superpartners, that however, as we discussed, are actually massive,
with a non-perturbative mass at the Planck scale).
These degrees of freedom are suitable to arrange in
two doublets of two different $SU(2)$ subgroups
of the symmetry group. Each doublet is therefore like 
the ``up'' and ``down'' of an $SU(2)$ doublet of the weak interactions,
but this time with a multiplicity
index 4. As we will see, this 4 will break into 3+1, the 3 
corresponding to the three quark flavors, and the 1 (the singlet) to
the lepton.  The number of degrees of freedom
seems to be the correct one, because, in order to build up
massive particles, we have to couple \emph{two} chiral fields,
left and right say. Their transformation properties seem however
to go in the wrong direction: in our case in fact all the fields
are charged under an $SU(2)$. If we make on the other hand
the hypothesis that for some reason the two $SU(2)$ are identified, 
we obtain that the weak interactions are not chiral,
in evident contradiction with the experiments.  
Although surprising, the spectrum we find is nevertheless correct; 
the point is that, as long as we don't consider the compactness of 
the string space-time, as we cannot
introduce masses for the light particles and fields,
we cannot even speak about breaking of parity in the weak interactions.
This is what we will discuss in the following section.

\noindent
\subsection{\sl Light masses}

Once compactness of space-time is taken into 
account, the most ($Z_2$--) singular configuration
of the system appears to be the one in which
orbifold operations act also on the non-twisted coordinates.
In this case, the only allowed operations left over are shifts.  
As we have seen, shifts in the ``internal'' coordinates lead to masses
that, owing to the fact that such coordinates are
also twisted, remain for ever stuck at the Planck scale. 
Of this kind is the ``shift'' in masses that breaks supersymmetry.
Further shifts in the space-time coordinates produce mass shifts that
``decay'' as they expand. Since the expansion of these coordinates 
is related to that of the time coordinate, this means that 
such masses decay with time. The precise relation
between the mass scales and the time scale is 
not simply the one we would expect by the Uncertainty Principle:
$m \sim {1 \over {\cal T}}$, the behavior we would expect
from a pure field theory analysis. The String Theory space is an ``orbifold'',
and the low string modes ``feel'' the space-time in a different way.
Since however the orbifold shifts lift  positively
\emph{all} masses inside supersymmetry multiplets,
the mass gap between particles and ``s-particles'' remains, for any time,
at the Planck scale.

Besides the conceptual problem of viewing as ``functions'' of space-time
what we are used to consider instead as ``stable'' parameters of our world, 
relating masses of  particles to the age of the Universe
seems to pose anyway the problem of explaining the GeV scale, 
namely a scale ``only'' $10^{-19}$ times less than the Planck mass but
still much higher than the inverse of the time scale.
If we make the rough approximation of considering the compactification space
as a product of circles, 
we would expect masses to be given by an expression like:
\be
m^D_{(0)} \, \approx \, \prod_{i=1}^{i=D}{1 \over R_i} \, ,
\label{m0}
\ee
where $R_i$ are the ``radii'' of the various coordinates. 
The index $i$ runs then from 1 to the
dimensionality $D$ of the theory (e.g. $D=11$ in the case of M-theory). 
In the (more realistic?) case of compactification on a curved space,
such as a sphere or an ellipsoid,
expression (\ref{m0}) then simply states that the mass scale is set by the
``curvature'':
\be
m_{(0)} \, \sim \, {1 \over {\cal R}} \, ,
\ee
with ${\cal R}^D =\prod_{i=1}^{i=D}R_i$.
Our problem is to understand what is precisely the ``correct geometry''.
Clearly, Since many coordinates are frozen at  the Planck scale, 
${\cal R}$ cannot be just the ``Age of the Universe''.
Moreover, we have to understand how different particles can ``live'' 
on different spaces, and therefore have different masses.

A first observation is that, owing to the fact that the string space is
an orbifold, the string ``age'', ${\cal T_{\rm string}}$,
does not coincide with what we perceive as the Age of the Universe,
${\cal T}$: any $Z_2$ shift implies in fact a square-root relation between
${\cal T_{\rm string}}$ and ${\cal T}$. In order to see this, 
consider the Uncertainty Principle on
such a $Z_2$ orbifold: owing to the $Z_2$
identification inside this space, the error $\Delta R$
around the edge of the space (or time) coordinate, set at ``$R$'', 
is half of the error of the torus. We have therefore:
\be
{\Delta R \over R} \vert_{\rm orbifold} \; = \; {1 \over 2}
{\Delta R \over R} \vert_{\rm flat~space} \, .
\label{deltar}
\ee 
Integrating both sides, we get: 
\be
\log R_{\rm orb.} \, = \, {1 \over 2} \log R_{\rm fl.\,sp.} \, ,
\ee
which precisely implies the square-root relation among the two coordinates
\footnote{Another way to see this is by considering that
a $Z_2$ projection introduces a factor ${1 \over 2}$ in
expressions like (\ref{mu}), that give the running of couplings
as a function of the space-size: 
$\beta \log \mu \, \to \, {1 \over 2} \beta \log \mu \,
= \, \beta \log \sqrt{\mu}$. This implies that the effective space-time
of the projected theory is the square-root of the unprojected one.}.
In our space-time there is room for two independent shifts,
plus a set of Wilson lines. We want now to see their effect
on the way the various particles feel space and time. 
We have seen that, before any shift in the space-time coordinates,
any twisted sector gives rise to
four chiral matter fermions transforming in the ${\bf 4}$
of a unitary gauge group.
The first $Z_2$-shift in the space-time breaks this symmetry, reducing 
the rank through a ``level doubling'' projection.
This implies that 1) half of the gauge group becomes massive; 2)
half of the matter becomes also massive.
The initial symmetry is therefore broken
to only one $SU(2)$ \footnote{This is the maximal non-Abelian group.
We will discuss in section \ref{photon}
about the fate of the $U(1)$ factors.}, 
under which only half of matter transforms.
The remaining matter degrees of freedom become massive. Since
this stringy phenomenon happens below the Planck scale, it admits 
a description in terms of field theory; 
from this, we see that, in order to build  up massive fields, these
degrees of freedom must combine with those that a priori were left
massless by this operation. Therefore, of the initial fourfold
degeneracy of (left-handed) massless matter, we make up light
massive matter, of which only the left-handed part feels the 
$SU(2)$ symmetry, namely what survives of the initial symmetry.
This group can be identified with the electroweak group 
\footnote{We skip here for simplicity the issue of the mixing with
the electromagnetic $U(1)$, of which we will talk later.}.
The chirality of weak interactions comes out therefore as a
consequence of a shift in space-time. This had to be expected:
the breaking of parity is in fact somehow like a free orbifold projection
on the space-time. 
It is legitimate to ask what is the scale of mass of the gauge bosons 
of the ``missing'' $SU(2)$, namely whether
there is a scale at which we should expect to observe an enhancement of 
symmetry. The answer is no, there is no such a scale.
The reason is that the scale of such bosons is simply T-dual, with 
respect to the Planck scale, to the scale of the masses
of particles. This fact can be understood by mapping
this problem to a specific string configuration realizing such effect:
in any string realization of such a situation, the gauge
bosons originate from the ``untwisted'' sector, while matter 
comes from the ``twisted'' sector \footnote{In order to visualize the
situation, it is convenient to think in terms of heterotic string.
Although a priori all the string constructions are equivalent, this
is the one in which this kind of phenomena are more explicitly 
described.}. It is clear therefore that a shift on the string lattice 
lifts  the masses of gauge bosons and of matter in a T-dual way.
Since the scale of particle masses is several orders of magnitude
below the Planck scale,
the mass of such bosons is much above the Planck scale; at such scale,
we are not anymore allowed to speak about ``gauge bosons'' or,
in general, fields, in the way we normally intend them.

A second $Z_2$ breaks then the symmetry inside each $SU(2)$ 
multiplet. This results in a non-vanishing mass of the corresponding
gauge bosons. We will discuss below how these operations precisely lead
to the ``standard'' spectrum of quarks and leptons, with the
unbroken electromagnetic and broken weak group of interactions.
For the moment, we are interested in seeing what is the mass scale
corresponding to the breaking of this second $SU(2) \approx SU(2)_{\rm w.i.}$.
For the gauge bosons of $SU(2)_{\rm w.i.}$ the situation is different
from the one of the first shift: in this case, 
the ``shift'' associated to the projection
that breaks the symmetry inside the multiplets cannot be a further 
level-doubling rank reduction: the only possibility is that it acts
in a way equivalent to a Wilson line.
As is known (see for instance ref. \cite{kk}), a Wilson line
acts as a shift in the windings, namely, as in the twisted sector.
This is why the electroweak scale is of the same order of the highest
particle's scale, that corresponds to the top-quark mass. 
Therefore, we expect this scale to roughly correspond 
to the quartic root of the inverse
of the Age of the Universe. This gives 
$\sim 10^{-15} {\rm M}_{\rm P}$, i.e. $m_{\rm e. w.} \approx {\cal O}
\left( 10^4 \, {\rm GeV}  \right) $, almost, but not quite exactly, 
corresponding to what actually it is. The point is that
our evaluation at this stage is too naive: for instance, 
with just the two operations on the space-time sofar considered,
we cannot distinguish
between the three sectors that give rise to the replication of matter
in three families.
Two shifts in the space-time directions are not enough to 
account for the spreading of masses as we observe it:
as long as we
consider the string space as factorized in ``space-time'' and
``internal space'', these sectors will always appear degenerate. 
In this naive set up,
the masses of our, ``observable'' world, would be 
related only to the scale of the extended coordinates, 
according to the following expression, the same for all particles:
\be
m \, \sim \, \left( {1 \over {\cal T}_{\rm str.}^4  } \right)^{1 \over 4} \,
\times \, \left( 1^{D-4} \right)^{1 \over D-4} \, .
\label{mf}
\ee
However, in this way the string space appears as a ``trivial''
bundle, with a base, the space-time, just multiplied by a fiber,
the internal space. All families appear on the same footing.
From a very rough point of view, this scheme 
is acceptable, as long as it is a matter of qualitatively explaining 
the difference between compact twisted, and extended, 
expanding coordinates, and why
observable particles have masses so small as compared to the Planck scale.
However, the condition of ``minimal initial entropy'' requires 
the introduction of more operations, that further ``curve''
and complicate the configuration of the space,
so that the internal space cannot be so simply ``factorized out''
from the external one. What is left is actually room
for Wilson lines. Generically, a Wilson line is an object that
acts on two pieces of the string space: it links an operation
on some charges of an ``internal space'' to the charges, or windings,
of other coordinates. In some sense, it provides therefore
an ``embedding'' of the internal space it acts on, onto the space
of the other coordinates. In our case, the Wilson lines under consideration
provide an ``embedding'' of the internal coordinates
in the external ones. We can roughly account for their effect by
saying that, at any time such a Wilson line effectively acts, 
one (or more) of the
$D-4$ internal coordinates, instead of contributing to the second
factor, has to be counted as contributing to the first factor
of the r.h.s. of (\ref{mf}). The result is that the mass expression
gets modified to something like:
\be
m \, \sim \, \left( {1 \over {\cal T}_{\rm str.}^4 \times 1^n}
\right)^{1 \over 4 + n/2} \,
\times \, \left( 1^{D-4-n} \right)^{1 \over D-4-n} \, ,
\label{mn}
\ee
where $n$ is the number of the embedded coordinates, $0 \leq n \leq D-4$
(we will explain below why only $n/2$ enters in the mass expression)
and for ${\cal T}_{\rm str.}$ we intend now $\sqrt{\cal T}$, the
square root of the age of the Universe.

Even without a detailed knowledge of what
precisely the most singular configuration is, we can already have a
qualitative insight by taking as usual the approach
of parametrizing singularities through $Z_2$ projections.
We can therefore think at such Wilson lines as a set of
$Z_2$ shifts, acting, as usual for Wilson lines, on the windings.
Namely, they generate a set of corresponding ``twisted sectors'',
from which the various particles arise.
As is known, whenever a $Z_2$ projection acts on the coordinates,
the effective size of the space is reduced by a factor two.
Correspondingly, the mass scale is doubled.
In the case of these Wilson lines, mass scales are instead
halved, because these projections act only in a ``T-dual'' way,
on the ``twisted sector''. 
Whenever such a Wilson line is active, namely,
whenever we look at the corresponding ``twisted sector'', 
the mass scale gets an extra factor 1/2.
Taking into account also the rescaling factors on the mass scales
due to the three ``true'' $Z_2$ orbifold twists of (\ref{zzz}),
we obtain the following spectrum of masses:
\be
m_n \, \sim \, 2^{3-n}
\left( {1 \over {\cal T}_{\rm str.}^4 \times 1^n} \right)^{1 \over 4 + n/2} \,
\times \, \left( 1^{D-4-n} \right)^{1 \over D-4-n} \, .
\label{mm}
\ee 
Notice that, while the number of rescaling factors due to the Wilson lines
varies from sector to sector, according to whether the corresponding
Wilson lines are or not active, the rescaling factor due to the 
true orbifold projections is always present. This corresponds to the
fact that these projections act on all the sectors.
We want here to remark that between internal and external
coordinates there is an essential difference 
also for what matters the effect produced by $Z_2$ 
shifts. As we have seen, when it acts on the internal coordinates,
any such shift produces a rescaling by a factor $1/2$. When it acts 
on the space-time coordinates, it
leads instead to a ``square root law'' for the proper time.
The reason of this behavior is that, when acting on untwisted coordinates,
such a shift produces a ``freely acting orbifold'', in which the
linear dependence on the string coordinates is suppressed to a
logarithmic behavior: $f(X) \sim X \,  \to  \, f(X) \sim \log X$.
In this case, 
a $1/2$ normalization factor for $f(x)$ is then traduced into a square root
rescaling for the string coordinate.
In the internal space, all coordinates are twisted, and although
untwisted sectors are also present, the dominant behavior is determined 
by the twisted sectors. In these latter, the twist superposes to, and therefore
``hides'', the shift.
The reason why only $n/2$, namely half of the number of embedded internal
coordinates enters in expressions (\ref{mn}), (\ref{mm}), is that, 
without Wilson lines, each (space-time transverse)
coordinate would be shifted only once, so that its ``proper length'' 
would be ${\cal T}_{\rm str.}$, the square-root of the Age of the 
Universe. Once we switch on a Wilson line, the space-time coordinate,
in the appropriate ``twisted sector'' from which 
the corresponding matter arises, feels a double shift, and therefore
follows a quartic root law.

Expression (\ref{mm}), derived on the base of ``dimensional''
considerations, helps to understand how a varied spectrum of masses
may arise. It however can by no means be considered to
give the precise expression for real masses. In order to get this,
we should better understand the details of the ``most singular
compactification''. Even without that, the above arguments allow us 
anyway to figure out, at least from a qualitative point of view,
how it is possible to arrange masses in a
(roughly) exponential sequel. More precisely, we are able to refer the
exponential arrangement of families 
to the increasing number of coordinates embedded
into the space-time for the different orbifold twisted sectors,
giving rise to matter replication.
Inside each twisted sector, the $SU(2)$-breaking shift
acts then as a ``square root law''. This on a different proper length
for each family. 
The mass of the gauge bosons corresponds to the breaking scale,
and is related to the highest mass~\footnote{We will come back to the 
relation between the mass of  
gauge bosons and matter states in section~\ref{massl}.}.  
This turns out to be of the order of:
\be
m_{\rm e.w.} \, \sim \, 2^{3-n} 10^{-30.5 \times {4 \over 4 +n/2}} \, 
{\rm M}_{\rm P}
~ \sim \, 2^{3-n} 10^{-30.5 \times {4 \over 4 +n/2}} \times 1,2 \times 10^{19} 
{\rm GeV} \, ,
\label{mew}
\ee
which gives the best fit for $n=8$: 
\be
m_{\rm e.w.}\,  \propto \, m_8 \, \approx \, 200 \, {\rm GeV} \, .
\ee
(For $n=7$, corresponding to $D=11$, we would get $m \approx 35$ GeV).
This fit strongly suggests that the number of internal dimensions
is 8, implying that the theory is originally twelve-dimensional.
However, it does not necessarily have to be supersymmetric.
We will come back to this point in section~\ref{comments}.

\subsection{\sl The low-energy spectrum}

We have seen how three matter generations 
arise in String Theory. We have also seen what is the scale 
corresponding to the heaviest observable 
particles/fields, and that this necessarily corresponds to the scale at which 
$SU(2)_{\rm w.i.}$ is broken. 
In this section we want to discuss what is the observable spectrum of 
String Theory. 
Namely, what are the particles and fields propagating at a scale
below the Planck scale.
We start by discussing what are the massless states,
namely, what are the states that remain \emph{strictly massless}.

\subsubsection{\sl untwisted vs. twisted sector}

As we have seen, all the states appearing on the string ``twisted sectors'',
therefore all the matter states, acquire a mass,
that, for each particle, depends on the corresponding (proper) 
space-time scale.  Namely, they would be strictly massless
if the observed space-time was infinitely extended \footnote{Remember
that, as required by minimal initial entropy,
each ``orbifold twisted sector'' has a ``rank four'' group
with (bi-charged) matter in the fundamental representation. Of 
the initial rank 16 per each such sector, a rank 12 part
had already been lifted to the Planck scale by ``rank-reducing'' projections,
acting as shifts on the ``internal'' coordinates.}. As we discussed, 
the less entropic situation, the one the theory
starts with in our branch,  is attained at time $t=1$. 
By approaching this point from a time $t < 1$, 
the next to the final step, before the minimal entropy is reached,
is, if we consider the non-twisted coordinates as infinitely extended,
a ${\cal N}_4=1$ vacuum. Once the last projection took place,
supersymmetry is not only reduced but necessarily completely broken.
This step may appear as perturbative or non-perturbative, depending on the
approach one chooses. Anyway, the masses of all the superpartners
are lifted to the Planck scale.  As seen from a  ``heterotic point of view'',
this breaking is non-perturbative,
corresponding to a ``shift'' along the  ``dilaton''
coordinate, which, as we have seen,
is stuck at the Planck scale.
The situations of gauge fields is reversed with respect to the one
of matter fields:
matter states appear in the ``twisted sectors'', while
gauge multiplets 
always show up in the ``untwisted sector'' of the theory. 
As a consequence, also
the supersymmetry-breaking shift acts in a reversed way, by lifting 
the mass of the fermionic states (gravitino, gauginos) instead of that
of the bosonic states, superpartners of the fermionic particles.
Therefore, at the point of minimal entropy,
the low energy world is made up of light fermionic matter
(\emph{no scalar} fields are anymore present!),
and of massless/massive gauge fields. Matter states appear
to be charged with respect to bi-fundamental representations.
More precisely, they transform in the $({\bf 4}, {\bf 4^{\prime}})$, 
replicated in three families: $({\bf 4}, {\bf 4^{\prime}})$,
$({\bf 4}, {\bf 4^{\prime \prime}})$, 
$({\bf 4}, {\bf 4^{\prime \prime \prime}})$.
We will discuss here how the ${\bf 4}$, common to all of them, is the
one which gives rise to the "$U(1)_{\rm e.m.}$"$ \otimes SU(3)_{\rm color}$,
while the ${\bf 4^{\prime}}$, ${\bf 4^{\prime \prime}}$
and  ${\bf 4^{\prime \prime \prime}}$ are all broken symmetries.
Roughly speaking, these correspond to "$SU(2)_{\rm w.i.}$"$ \times SU(2)$
(the ``~'' symbols are there to recall that the actual $U(1)_{\rm e.m.}$
and $SU(2)_{\rm w.i.}$ are mixtures of the two).
The fact that the electroweak symmetry appears to be replicated in three
copies is only an artifact of our representation of the states.
Indeed, for the string point of view, there is no such a symmetry, 
being all the corresponding bosons massive. The String Theory vacuum
\footnote{From now on, we will refer to the configuration deriving from
minimization of entropy at $t=1$ as to ``the'' String vacuum.}
knows only about the ``$({\bf 4}, \ldots )$''. In Ref.~\cite{striality}
we discussed how, when there is matter bi-charged under two gauge groups,
originating from two sectors non-perturbative with respect to each other,
like are here the ${\bf 4}$ and ${\bf 4^{\prime}}$,
${\bf 4^{\prime \prime}}$, ${\bf 4^{\prime \prime \prime}}$, then,
whenever supersymmetry is reduced to ${\cal N}_4=1$, either the one
or the other sector/group is at the strong coupling. 
This means that, in a string framework, talking about gauge symmetries 
as one usually does in a field theory approach is not much appropriate:
these symmetries are at least in part already broken, and therefore
do not appear in the string vacuum. 
More than talking about gauge symmetries, what we can do is therefore
to just \emph{count} the matter states/degrees
of freedom, and \emph{interpret} their multiplicities as due to their 
symmetry transformation properties, 
as they would appear, in a field theory description,
from  ``above'', before flowing to the strong coupling.
This is an artificial, pictorial situation, that does not
exist in the actual string realization. That's fine: the string vacuum
indeed describes the world as we observe it, namely, with quarks
at the strong coupling. If we look at this configuration
from the point of view of the ${\bf 4}$, we have a description
of an ``unstable'' vacuum, in which $SU(3)$ has yet to flow to the
strong coupling, and we observe that the ``quarks'' have multiplicities
${\bf 4^{\prime}}$, ${\bf 4^{\prime \prime}}$, 
${\bf 4^{\prime \prime \prime}}$, corresponding to the three twisted sectors
they belong to. On the other hand, had we observed the story from 
the reversed point of view, namely by looking from the point of view
of the weakly coupled sector, then we would have said that all
the quarks are charged under \emph{one} common weakly coupled gauge group,
and under three ``hidden'' gauge groups at the strong coupling. 
Namely, what we count is indeed
the number of matter states; we are not allowed to count gauge bosons
of broken, or strongly coupled, symmetries. So, either we chose
to look from the ``weakly coupled $SU(3)$'' point of view, and we have
the fake impression that $SU(2)_{\rm w.i.}$ is triplicated, with
an independent $SU(2)_{\rm w.i.}$ per each family, or we look from the
point of view of $SU(2)_{\rm w.i.}$; in the second case we count only
one such group and an artificial triplication of $SU(3)$. 
I stress however that
for the string, both these symmetries do not exist: they are only a guideline
for our convenience. Notice also that the heterotic realization
of the vacuum corresponds to an unstable phase in which the ${\bf 4}$
is realized on the currents. The Wilson line the breaks it further
to $SU(3)$ is non-perturbative. The fact that there is no
explicit string realization of the vacuum corresponding to our world,
in which we can follow the path of single quarks,
is somehow unavoidable: it is in fact related to the fact that
at low energy only $SU(3)$ singlets exist as free, 
asymptotic states in Nature.

\subsubsection{\sl the splitting of the ``untwisted'' gauge group:
origin of the photon and ``$SU(3)$''}
\label{photon}

We have seen that in the string vacuum
at most the ${\bf 4}$ can remain massless.
The legitimate question is now whether also this is at the end of the day
lifted by some shifts. The answer is that
there is no room to accommodate further, independent
shifts, able to reduce the rank of this sector: all the internal coordinates
have been already used, in order to reduce the rank of the
``twisted'' sectors to ``${\bf 4}^{\prime}$'', and the space-time
transverse ones to break further this group and its replicas,
${\bf 4}^{\prime \prime}$, ${\bf 4}^{\prime \prime \prime}$,
down to the observable, broken weak symmetries of our world.
As we already remarked,  
since at least one coordinate of the
theory is extended, the number of allowed orbifold operations can be 
analyzed within the framework of a perturbative string orbifold.
Elementary properties of string orbifolds tell us
that with this we exhausted all the possible
``rank reducing shifts'', and the only possible operation on the
${\bf 4}$ is a further breaking driven by Wilson lines, the
ones that distinguish the matter states transforming in this
representation into different mass levels \footnote{Notice that,
in a heterotic-like vacuum in which the ${\bf 4}$ is realized
on the currents, the double rank reduction of this sector
is realized, as observed in Ref.~\cite{striality}, by shifts that
act always, besides the ``observable'' compact space,
also on the ``dilaton plane'', a non-perturbative plane
for the heterotic string. In dual constructions this plane appears
as a normal, two-dimensional orbifold plane, in which it is easy to see
that only two independent shifts can be accommodated.
These correspond to a maximal double rank reduction, from 16 to 4.}. 
From: i) the counting of
the matter degrees of freedom, ii) knowing that this sector
is at the strong coupling as soon as we identify the 
``${\bf 4}^{\prime}$'' as the sectors containing the group of
weak interactions, and iii) the requirement of consistency with
a field theory interpretation of the string states below the Planck 
scale~\footnote{for us, field theory is not something realized
below the string scale. It is rather an approximation (that locally works),
obtained by artificially considering the space-time coordinates as
infinitely extended. As a consequence, the ``field theory'' scale
is in our set up the scale of the compact, but non-twisted, string coordinates.
In order to keep contact with the usual approach and technology,
what we have to do is to send to infinity the size of space-time,
and consider the \underline{massless} spectrum of four
dimensional string vacua.},
we derive that the only possibility is that this representation
is broken to $U(1) \times SU(3)$. A further breaking would in fact
lead to a positive beta-function, in contradiction with point ii).
Actually, point iii) tells also us that $U(1)$  
combines in some way with  part of the twisted sector's gauge
group to  build a non-anomalous $U(1)$, the only allowed to survive.
This gives rise to the photon. Leptons and quarks originate
from the splitting of the ${\bf 4}$ of $U(4)$ as ${\bf 3} \oplus {\bf 1}$
of $SU(3)$.

\subsubsection{\sl about the Higgs field}
\label{higgs}

If the origin of masses is a purely Stringy phenomenon, what about the
Higgs field? The point is simply that in String Theory
there is no need of introducing such a field:
its ``raison d'\^{e}tre'' was justified in a field theory context, 
in which, once realized the renormalizability of gauge theories, one advocates
the principle of ``spontaneous breaking'' of such a symmetry: in spontaneously
broken gauge theories renormalizability is then ensured.
But here, renormalizability of the theory holds just because it is 
string theory, which is not only renormalizable but finite.
Masses are the consequence of the ``microscopic'' singularity of 
the string space. 
Precisely thanks to the existence of masses, the microscopic
singularity generates in turn space-time singularities. 
String Theory provides therefore  a sort of 
``micro/macro unification'': 
microscopic phenomena are mirrored by macroscopic ones.

Even though the origin of masses does not rely on the existence of
a Higgs field, something of the old idea of Higgs particle 
and potential nevertheless survives, although in another form.
It is known that, in non-supersymmetric string vacua,
at certain points in the moduli space tachyons appear \cite{adk}.
Such tachyons play precisely the role of Higgs fields. Namely, they have a
``mass'' term with the wrong sign. Moreover, since
the first massive state to come
down to negative mass as we move out
from the physical region of parameters
is of course the ``superpartner'' of the lightest
observed particle \footnote{Remember that all masses inside
multiplets of the broken supersymmetry are positively lifted of the
same amount by a space-time shift.}, we can think at it as a field
that, coming from a multiplet of broken supersymmetry, feels also
the interactions of the observed particle. In particular, 
if we represent these degrees of freedom and their couplings
in terms of fields, we see that cubic and
quartic terms are automatically generated by interactions.   
All that to say that the ``effective action'' for such a state
is precisely a Higgs-like mass plus potential term.
The physical vacuum of string theory is the one in which this tachyon
becomes massive.  It is therefore not anymore perceived as a tachyon, but
rather as a massive state, with a mass at the Planck scale. However, the fact
that it comes from a Higgs-like state tells also us that its ``superpartner'',
the lightest observed particle, must be electrically neutral. Otherwise,
also $U(1)_{\rm e.m.}$ would be broken, with a massive photon, something we
have seen it does not happen in the  vacuum.
The idea of the Higgs field as the superpartner of an observed particle
was already considered in the past. It was discarded because of the
phenomenological non viability of certain interactions it would
then had. However, here we are not stating that
the Higgs is the superpartner of the neutrino. We are saying that
the Higgs \emph{does not exist} at all 
\footnote{or, if one prefers, 
it exists but its mass is at the Planck scale, therefore it can be 
considered as decoupled.}. 
It is just the interpretation of
certain string states in an unphysical region that resembles 
what we have in mind for the Higgs. And precisely this technical
fact forces the lightest observed particle to be electrically
neutral, namely, the neutrino to exist.

\subsubsection{\sl mass levels}
\label{massl}

Once realized that the lightest particle must be electrically neutral,
all other charge assignments follow automatically from the
requirement of anomaly cancellation of the light spectrum, the one that
admits a representation in terms of field theory~\footnote{Notice
that the correspondence with field theory is stated at the 
\underline{massless} level, for massless fields and particles.
As we have seen, the inclusion of masses goes beyond the domain of field
theory.}. The 
particle spectrum of the Standard Model is therefore automatically
implied by  ``minimal initial entropy'', 
and by the finiteness, at any time of observation, of our horizon. 
Owing to the ``replication'' of the spectrum
into three families, consequence of the replication of orbifold sectors,
we expect the above described ``Higgs condensation'' argument
to be valid for the three families. Therefore,
we expect the three neutral particles to be the lightest ones.
According to (\ref{mm}), the  
lightest mass, which should correspond to the lightest neutrino, is:
\be
m_0 \, \sim \, 2^3 \times 10^{-30.5} \times 1,2 \times 10^{19} \, {\rm GeV} 
~~ \sim 3 \times 10^{-1} \, {\rm eV} \,
\ee
which is below the experimental bound for the mass of the electron's neutrino.
The following two mass steps, according to (\ref{mm}), are:
$m_1 \, \sim \, 2^2  \times 10^{-30.5 \times 8/9} \, {\rm M}_{\rm P} \, \sim
\, 30  \, {\rm eV}$ and
$m_2 \, \sim \, 2 \times 10^{-30.5 \times 4/5} \, {\rm M}_{\rm P}
\, \sim \, 10 \, {\rm KeV}$, which are also below the actual experimental
bounds for $\nu_{\mu}$ and $\nu_{\tau}$.
The fourth mass step, that we expect to correspond to the first
excitation of charged matter, is
$m_3 \, \sim \, 10^{-30.5 \times 8/11} \, {\rm M}_{\rm P} \, \sim \,
0.55 \, {\rm MeV}$, quite close to the electron's mass.
This agreement, although not simply a matter of coincidence, has on
the other hand to be taken as just an indication. The mass
steps contained in expression (\ref{mm}) are in fact not enough
to account for the difference between all particles and the $W$
and $Z$ bosons. This means that our way of parameterizing
the singularity of the space is too simple, and has to be taken
only as a rough approximation, more suitable for a qualitative
than for a quantitative analysis. For instance, (\ref{mm})
does not account for the effect of
all the ``Wilson lines'' present in the theory. We know in fact that
all the three ``twisted sectors'' are distinguished from each other,
and that, inside each of them, Wilson lines distinguish matter by
breaking the $SU(2)$ multiplets, and the ${\bf 4}$ into ${\bf 3} + {\bf 1}$.
However, by just counting the coordinates we see only nine steps.
What is lacking, and we would need, is a description of the 
``non-linear'' effects produced by all these operations, 
which most probably require a treatment of the 
string space in which the coordinates are not simply factorized.
Another strong reason why
our arguments have to be considered rather qualitative is that,
from a rigorous point of view, it is not possible to 
construct a perturbative ``$Z_2$ string orbifold'' 
in which all the operations we have mentioned are explicitly realized.  
Our discussion is just intended
to give the flavor of what is likely to happen in a curved geometry,
a sort of ``ellipsoid'', in which the coordinates,
not factorizable, are in part very large
and in part of length one.

Even with our big simplifications, we can however understand
some mass relations, whose explanation  is due to pure stringy  
effects. For instance, String Theory tells us \emph{why} the mass of the
$SU(2)_{\rm w.i.}$ bosons, $(W,Z)$, is roughly the average between
the top and bottom masses: $m_{W,Z} \approx {m_t + m_b \over 2}$. 
In order to see this, let's analyze more in detail
the structure of matter states. As we said, each family appears
from states that, were not for the compactness of space-time, would be
strictly massless, and transform in a $({\bf 4},{\bf 4^{\prime}})$,
bi-fundamental representation. ${\bf 4^{\prime}}$ is then broken by
shifts in the space-time, while the ${\bf 4}$, as we discussed,
gives rise to the photon and the strong interactions.
Let's concentrate on the ${\bf 4^{\prime}}$.
In the following, we will look at the problem by either
1) considering matter as appearing in the twisted sector
of a heterotic realization; in this case the
${\bf 4^{\prime}}$ will not appear explicitly, but only through
its discrete subgroup of the multiplicity of states in the twisted sectors
\footnote{This is the point of view we used 
when discussing the origin of the photon and the gluons.}, 
or 2) mapping the problem 
into a description in which the ${\bf 4^{\prime}}$ 
appears in the currents of a heterotic construction.
For reasons that we will explain in a moment,
these two representations are equivalent,
and the choice of 1) or 2) is dictated by convenience, depending on
which phenomena we want to observe explicitly.
Let's start with the picture 1): the string configuration with minimal
entropy requires the action of a double shift in space-time.
As we have seen, a consequence of this is the breaking of Parity at the
Planck scale \footnote{More precisely, above it.}, and 
the breaking of the surviving
$SU(2)_{\rm w.i.}$ by a ``$Z_2$-Wilson line'', that
splits masses inside matter doublets. This is the stringy realization of the
Higgs phenomenon. As discussed above,
the order of magnitude of such Wilson line essentially sets 
the ``top-quark mass''. In order to better investigate
how matter and boson masses are related, we can pass to picture 2),
namely, we map the problem to a description in which both matter and
gauge states are realized as part of the currents of a heterotic vacuum. 
This is possible because Wilson lines act on the windings, 
treating gauge bosons
as if they had the same origin as matter, on the ``twisted sector''.
Therefore, the states that
in picture 1) appeared in the twisted sector, matter and ``inexplicit'' gauge,
in picture 2) appear in the currents, with the quarks corresponding to
matter in the untwisted sector. 
In this second picture, matter transforms in the fundamental of a gauge group,
with an internal index (the ``color index'')
on a ``twisted sector'' gauge group. Under orbifold shifts, it receives
therefore the same kind of mass lift as the gauge bosons.
In this representation,  the situation corresponding to our
physical problem appears as follows:
one starts with a symmetry, realized now in the untwisted sector, whose
gauge bosons arise as states of the kind $A_{\mu} =
\overline{X}_{\mu} \otimes \phi_i \phi_j$, in terms of left and
right moving vertex operators. 
Let's restrict to the case of four possible values
for $i$ and $j$. In this case we have a symmetry $U(2)$,
that we can consider as the one surviving after the first shift
in the space-time. 
For simplicity, let's neglect here the fact that
in the physical situation the anomalous
$U(2)$ is  broken to a non-anomalous $SU(2)_{\rm w.i.}$.
Since we want to consider the breaking of the ${\bf 2}$, 
let's group the four indices into two groups, 1 and 2.
Our $W$ bosons are represented in terms of vertex operators by
$\overline{X}_{\mu} \otimes \phi_1 \phi_2$ and its conjugate. 
Now consider a shift (the ``Wilson line'' under consideration)
that lifts the mass of the oscillators with index $2$, 
with the property that all the oscillators 
$\phi_2$ are eigenvectors of the ``right moving''
energy-momentum tensor with mass ${\it m}$,
while the non-shifted states, $\phi_1$, remain massless.
The mass of extra-diagonal states, such as the $W$ bosons,
and of linear combinations of ``diagonal'' bosons, 
$\overline{X}_{\mu} \otimes {1 \over 2 } \left( \phi_1 \phi_1
\oplus  \phi_2 \phi_2  \right)$, results then to be half of the mass of 
the states built entirely of oscillators with index ``2''.
Since we are considering a symmetry breaking due to
a Wilson lines that splits the mass inside matter doublets, and not
a lift in mass due to an ordinary rank reducing shift,
the mass of such vectors, $2 {\it m}$, is the same as the mass of the
lifted matter states, namely of the quarks ``up'' (in our case, the
heaviest one, the top quark). 
Therefore, $m_{W,Z} \sim {1 \over 2} m_t$. However, also the ``down'' quarks
are massive. All masses are then further shifted, and this relation    
is shifted to:
\be
m_{W,Z} \sim {1 \over 2} \left( m_t + m_b \right) \, .
\ee

\subsection{\sl Time-dependent masses?}
\label{masses}

Although very approximate, the above qualitative analysis
confirms that the relation between electroweak masses
and the time scale of the Universe we argued is essentially correct.
Since we are used to think at masses
as at ``stable'' parameters, certainly not quantities that
vary with time, this is somehow surprising. 
However, it is not  at all unreasonable.
It is in fact quite natural to think that the Universe
knows only about one fundamental mass/length scale, the Planck scale. 
From this point of view, all other scales appears as pure
``accidents'', related to the actual point in the evolution
of the Universe we live at (and make measurements).
The reason why we are used to think at all mass scales as constant,
is that their variation with time is quite slow. 
When we insert a mass parameter in an effective action
(like, e.g., the Standard Model Lagrangian), we are always thinking in 
terms of an infinitely extended space time. 
As long as we are only interested in ``local'' phenomena,
this assumption is reasonable, because the actual age of the 
Universe is big enough to allow
approximating our finite-size horizon with an infinite space-time.
However, when we push our field of investigation to the borders
of space-time, or of the domain of energy/momenta,
we cannot anymore neglect the ``boundary effects''
due to the fact that the portion
of Universe we can measure is finite. 
Considering the space-time as infinite corresponds
to artificially removing the natural cut-off. This produces the
infinities, and the (artificial) distinction between ``bare''
mass, which is divergent, ``physical'' mass, finite, and also
``running'' mass, related to the cut-off by the renormalization procedure. 
In some sense, here we have only a ``running mass'', although
this running has to be intended in a string/cosmological framework. 
Actually, the introduction of the horizon
as a natural cut-off would lead to a time-dependence of masses
even without considering the output
of the string theory analysis.
Indeed, what we are here proposing is that masses are a consequence
of the existence of this cut-off. In some sense, they ``measure''
the size of the Universe. From a purely field theoretical point of view,
mass terms are naturally created, from  ``nothing'', for all
particles and fields whose (zero) mass is not protected by some symmetry.
In this case, only the photon~\footnote{and the gluons} 
has a protected vanishing mass.
Since masses are not a consequence of ``contact'', or interaction, with
heavier particles or fields (no Grand Unification! \footnote{``Unification''
takes place at the Planck scale, where we are not anymore allowed to
speak in terms of fields and elementary particles.}), we don't have
to worry about hierarchy problems. The electroweak scale is 
trivially protected because it is the \emph{only} scale of the effective
theory. 
Remarkably, what string theory tells us, is that the time/cut-off dependence 
of masses is  different for different types of particles. 
In some way, they ``live'' on space-times with different scales 
\footnote{Taking the argument the other way round, we could say that
the fact that there exist different elementary particle's masses,
is a signal of the existence of internal dimensions.}.  
Namely, each type of particle has its own ``Uncertainty Principle'' relation,
leading to a mass related to the inverse of its ``proper'' time.
Although unexpected this may seem, the idea 
that different particles ``live'' on different spaces, namely, that
they feel in a different way space and time, is after all not that new in
Physics: the difference between Bosons and Fermions, namely between
particles with integer and half integer spin, is an example.
In that case we have in fact that, despite our macroscopic
perception of space and rotations inside it, elementary particles
live on a more complicated space.
We have seen that the space of String Theory is somehow an ``orbifold''
of what we call space-time. Different particles live on different
orbifold sectors, and they have each one a specific perception
of it.  It is difficult, if not impossible, to give an explanation
of these facts in terms of ordinary, four-dimensional field theory:
masses are additional degrees of freedom that, being sources of, and
charges of, gravity, only in a String Theory framework can be
consistently treated. They can well be introduced in a field theory 
framework, but only at the level of an effective theory, in which they 
appear as ``ad hoc'', external inputs \footnote{Even when they are introduced
through a Higgs mechanism, still they remain external inputs: simply the
problem is rephrased into that of explaining the Yukawa couplings
of the Higgs field.}, and at the price of 
neglecting an entire part of the quantum theory/interactions
they feel, i.e. gravity. 
This behavior of masses goes anyway so much in the opposite direction
of our common belief that it deserves some ``physical'' explanation.
Therefore, let's try to see whether it is possible to make sense of it
also with our intuition. Indeed, 
this phenomenon can be seen to be related
to the analogous behavior of the cosmological constant.
Our most common way of intending masses is through the concept
of ``inertia''. The mass of a particle tells us how much energy we have to 
spend in order to accelerate it of a certain amount. 
The existence of a cosmological constant tells us that
the Universe has a non-vanishing vacuum energy. 
Since energy, as masses, is source of gravity,
this means that in some
sense the background in which particles  ``live'' is not flat, but 
experiences a ``background gravitational force''.
Accelerating a particle is therefore equivalent to move it from
a geodesic to another one. If the cosmological constant decreases with time,
this means that as time goes by, the less and less is the energy
we have to spend to accelerate the particle, because we have to work
against a smaller ``background force''. Therefore,
what in practice we observe is that
the inertia of the particle decreases with time. 
We must however stress that this is only a ``rough'' argument, 
not able to make sense of the fact that different particles 
have a different time dependence. The true explanation comes 
from String Theory, and goes beyond our ``classical'' intuition
\footnote{Our ``intuition'' is based on the observations we make, 
and is related to the perception of space-time of photons and gravitons, 
which behave as ``truly'', ``elementary'' four dimensional objects.}.

Before discussing the implications of the time-dependence of masses, 
let's first stress that, as it is predicted
by string theory, at the present day
the rate of change of masses with time is very small.
Namely, the experimental uncertainty of mass values, $\Delta m / m$,
is larger than the expected change in such values due to the time
evolution, during the period since we talk about high-energy physics
and related experiments. Therefore,
this effect is not ruled out by today's direct experimental informations
about particle's masses.  
According to the string theory analysis,
during the evolution of the Universe, the particles, that 
at the Planck scale started all with a mass of the same order (the Planck
mass is the minimal mass allowed at such scale by the Uncertainty
Principle), become progressively lighter. This process is not uniform:
any type of particle behaves differently, as required by
the principle of minimal initial entropy. i.e. of maximal
initial differentiation. Asymptotically, all particles
will anyway meet at infinity, where all their masses will vanish,
as it will the average energy of the Universe per unit of volume.

Although not directly detected by current experiments, the decrease
of mass of particles can be traced in the very fact that the 
Universe expands \footnote{Another indirect confirmation
is provided by the analysis of the emission spectra of quasars \cite{ivanchik}.
We devote Ref.~\cite{alpha} to this issue.}. 
The fact that masses decrease is a consequence of the
expansion of the horizon, something a priori independent from the
expansion of the Universe itself. However, the fact that particles become
lighter implies a decrease of the attractive force, and an increase 
of the kinetic energy. This leads to an accelerated expansion,
with a decreasing rate of acceleration.
In order to see this, let's consider a simple ``model'' of the Universe.
We can make the
hypothesis that, at the Planck scale, the Universe, or better, 
the portion of Universe relevant for us, i.e. the one enclosed in
a ball with radius one in Planck units, is at ``equilibrium''.
By analogy with solid matter, we can imagine that the gravitational
attractive force is balanced, say, by some contact, repulsive ``electric'' 
force, that forbids all matter to collapse into a point
(the details of this force are not important, this is just an artifact 
to help fixing the ideas on something concrete).
The expansion of the horizon, due
to the existence of massless modes propagating in the 
``space-time'' coordinates, is equivalent to a perturbation that 
leads to a decrease of masses during time evolution. Owing to this,
all particles (galaxies) feel a nett repulsive force, due to 
a decrease of gravitational attraction. Let's consider
a particle (a galaxy) of mass $m$ located on the hypersurface at the 
border of the ball of radius $r$. This would experience a force given by:
\be
F \, = \, {m M \over r^2} - {q Q \over r^n} \, = \, m \ddot{r} , 
~~~~~ n > 2 \, ,
\ee
where $M$ stays for the mass enclosed in the ball, $Q$ 
its nett charge and $q$ the charge of the particle.
The ``initial condition'' is $F =0$ at $t=r=1$. 
For $t > 1$, it would move according to:
\be
\ddot{r} \, = \, {1 \over r^2} \left( {q Q \over m r^{n-2}} - M \right) \,
\propto  \,  {1 \over r^2} 
\left(  {1 \over t^q}  - {t^q \over r^{n-2}} \right) \, ,
\label{ev}
\ee  
where we choose an ``average'' time-dependence for masses, $m \sim 
{1 \over t^q}$ (for the lowest mass, $q = {1 \over 2}$).
One can easily realize that the motion we obtain,
namely the expansion of the ball, is accelerated, and the acceleration
decreases with time. 
For $r >> 1$, $t >> 1$, the dominant piece on the r.h.s. is
the first term in the brackets, because the ``contact force''
decays more rapidly (it is reasonable to assume that
$n$ is sufficiently large). Therefore, although the acceleration is positive,
its derivative is negative (remember that $q ~ \gsim ~ {1 \over 2}$).
For an ``average'' time-dependence of  masses, $\sim t^{q = 1 / 3}$, we can
solve (\ref{ev}) with the Ansatz $r \approx t^p$, obtaining $p = {5 \over 9}$,
a value quite close to $r \sim t^{2 / 3}$, the behavior expected 
from the usual estimates for a matter dominated Universe.
This also implies that the red shift due to the expansion rate, 
$ z_{\rm r. s.} \approx { \Delta R \over R} \sim {2 \over 3} t^{-1}$, 
dominates over the violet-shift 
$z_{\rm v.s.} \sim {\Delta m \over m} \approx - {1 \over 3} t^{-1}$,
induced by the negative derivative of mass, that raises all
atomic energy levels. What we observe is therefore still a red-shift.
At this point, it should be clear that the introduction of the
``balancing'' force $F \sim {q Q \over r^n}$ was just an artifact.
The true point is the increase of ``repulsive'' force, or, better,
of kinetic energy, due to the decrease of masses \footnote{Although evidence of
the acceleration in the rate of expansion of the Universe is 
provided by astronomical observations \cite{r,p}, the elaboration
of data and their exact interpretation is somehow model dependent.
At the present stage, we consider premature, in the framework of our proposal,
a detailed discussion of issues like the possible
very early era of ``decelerated expansion'', something suggested
for instance in Ref. \cite{tr} (just to give an example,
it could be that this effect disappears once the appropriate rescaling
of masses and emission spectra is taken into account).}.

\subsection*{\sl $P$ and $T$ parities violation.}

As we have seen, left-handiness of the weak interactions is a consequence
of the stringy origin of masses of particles. Namely, only left handed
states feel the weak interactions gauge group, while their right handed
counterparts would feel a group which is however broken at the (above the)
Planck scale. Electromagnetic interactions are on the other hand 
non chiral: the photon arises in fact as the non-anomalous, surviving massless
combination of the original gauge group bosons, under which transform
both the ``left'' and ``left-conjugate = right'' degrees of freedom,
corresponding to the ${\bf 2}$ and ${\bf 2}^{\prime}$ of the ${\bf 4}$.  
The breaking of parity results therefore as the effect of
shifts in the space-time. This had to be expected: parity 
is in fact defined as the symmetry under space reflections:
$x \to - x$. If the coordinate $x$ is shifted by a certain amount, 
$a: x \to x^{\prime} = x + a$, 
under $x \to -x$ we obtain  
$x^{\prime} \to x^{\prime \prime} = -x + a$. If the space was
infinitely extended, this shift would have no effect.
However, since the space is compact, it is not invariant under
translations, and parity gets broken. As we have seen, 
the order of magnitude of the parity breaking is T-dual,
with respect to the Planck scale, to that of
the shift under consideration, related to the
electroweak scale.

Finiteness of the horizon implies on the other hand the breaking
of the symmetry under time reversal. The reason is that
it introduces an energy gap, measured by the cosmological constant,
equivalent to a shift in the time coordinate. We expect therefore
the order of breaking of this symmetry to be  of the order of 
the cosmological constant.
This for what matters \emph{direct} violations of time reversal symmetry.
However, as is known, violations of this symmetry can be detected
also in an indirect way, through mixings in meson decays.
The ``CP'' violation parameters are then referable to a phase
in the mass matrix. This means that this effect is of the same order of the 
electro-weak scale. The reason is that not only the cosmological
constant, but also the masses of particles = energies at zero momentum,
are variables conjugate to time. Their introduction is therefore
also somehow equivalent to a shift in time, the ``specific'' time
of each kind of particle. A noteworthy peculiarity of String Theory
is that the breaking of time and space parities, 
while not immediately related in field theory,
are here instead two aspects of the same phenomenon:
as it can be seen by passing through an
Euclidean intermediate step, shifts in the compact space or time
coordinates are essentially equivalent, and interchangeable.

\subsection{\sl Gauge couplings and unification}

Gauge couplings receive 
contributions both from ``internal'' and ``space-time'' moduli.
We have seen that the ``internal'' moduli, that we generically indicate
by $X$, are bound at the Planck scale; their contribution,
approximately given as usual by functions of the type:
\be 
{1 \over g^2}(X) \, \approx  \, \log \Im X | F (X)|^4 \, + 
\, {\cal O} \left( {\rm e}^{-\Im X} \right) \, , 
\label{gx}
\ee
with $\Im X \sim 1$, $| F (1)|^4 \sim 1$, is therefore of order 1.
This is the order of the couplings at the ``unification'' scale.
What then distinguishes among the different couplings is their running below
the Planck scale, namely the part that depends on the space-time scale.
This is accounted for by the so-called ``infrared'' logarithmic
behavior, encoded in the term:
\be
\beta \log \mu \, ,
\label{mu}
\ee
where $\mu$ \footnote{We always work in Planck units, for which 
$M_{\rm Planck}=1$. Therefore, $\mu=1$ at the Planck scale.}, 
traditionally the infrared cut-off \cite{kk}, 
is precisely related to the space-time scale. 
In string theory, this term is computed through
a regularization procedure, according to which
$\mu$ is the inverse of the ``curvature'' of our extended space
\cite{kk}. Namely, it comes out by
artificially compactifying the physical space on a conformal
background roughly equivalent to a sphere.
This approximation, whose choice was dictated in Ref. \cite{kk}
by technical convenience, although not exactly
corresponding to the actual configuration of our space-time,
it captures nevertheless a fundamental feature, 
namely the fact that the cut-off is related
to the scale, or ``size'', of space-time, as are fields and particles masses.
In order to really discuss the running of couplings,
a full control of the non-perturbative corrections would be in order:
as we have seen, our Universe corresponds in fact to a strongly coupled
string vacuum. From a qualitative point of view,
we must however observe that, since the string approach which is the
closest to the description of the actual configuration 
is a perturbatively ${\cal N}_4=1$ vacuum, the running of the gauge couplings,
unless non-perturbative effects deeply affect the term \ref{mu}, 
may well be the one of a supersymmetric vacuum. Therefore, a heavy
breaking of supersymmetry, at the Planck scale, may be compatible
with the ``nice'' rescaling of gauge couplings allowing them to meet
\cite{drw}.
We must however stress that, being these considerations very qualitative,
one should look at these arguments more as giving the ``flavor''
of a possible explanation, than as providing a framework in which to
perform detailed computations and matching numbers. 
This is why we have been very
``floppy'', and in (\ref{gx}) we omitted normalization and integration 
constants. In a quantitative analysis, these cannot be neglected.
Here however their inclusion would not change the approximate
nature of this discussion.

\section{\bf Comments}
\label{comments}

We have seen how assuming a (thermo)-dynamical principle, that of
Entropy, as the dynamical ``input'' of string theory, leads
to a consistent picture of its evolution, able to justify
both the ``macroscopic'', cosmological evolution of our Universe,
and the ``microscopic'' details of space-time, namely,
the spectrum of elementary particles and their interactions. 
An essential point for this is the ``physical'' assumption that a theory of our
world \underline{at the present time} 
must be self-contained within our horizon of observation, 
namely, that it is not supposed to describe phenomena causally disconnected,
or belong to our "future".  This implies that 
String Theory, in which time is not an external parameter, but one of the
coordinates of the theory, must always be considered as
moving, at any time of its evolution, in a compact space-time.
Once implemented with this ``phenomenological principle'',
the ``thermodynamical principle'' can be seen to imply
also the Heisenberg's Uncertainty Principle. This appears first
at the cosmological level, through the existence of the
so called ``cosmological constant''.  
The dynamical principle contains 
in his ``program'' also the breaking of T-duality and 
space- and time-reversal symmetries,
something that makes microscopic and macroscopic phenomena
to be not exactly the same, and allows us to select an arrow
in the time evolution/space expansion.
Another dynamical effect is the separation of the string space
into  ``internal'' and ``extended'' one. The internal space is however
not simply factorized out. This leads to 
an intriguing relation between the curvature of the ``internal'' space, 
at whose singularities light matter appears, and the curvature of the 
``space-time'', generated by such a matter: massive
objects are ``singularities'' that generate the space-time curvature.
Unfortunately, here we have been able to present only a
rather qualitative analysis of these relations.
In order to go further and obtain results more precise than those
presented in this work, it is absolutely necessary to 
make computations in a well defined, mathematical scheme.
Although over-simplified, our analysis strongly supports
the hypothesis that the fundamental theory is essentially
twelve-dimensional; 
in order to embed the problem into a precise mathematical
framework, we would need 
a deeper understanding of this basic theory. In absence of that,
we can try at least to collect some informations ``from below''.
What we can learn from our analysis, is that there is no evidence
that  this higher dimensional theory should be supersymmetric.
All what we know is also compatible with the interpretation of the
twelfth coordinate as the order parameter for the breaking
of supersymmetry. The only fact that we know are that
1) when we can go to the limit $R_{12} \to 0$,
we obtain the so-called ``M-theory'', which presents supersymmetry;
2) once the theory has attained the configuration of maximal twisting,
supersymmetry is necessarily broken.
The twelve dimensional theory may be supersymmetric, but it could also be
that it is not, and that
M-theory is the first term in an expansion
of a non-supersymmetric theory around a supersymmetric vacuum.
It would look then supersymmetric
only because precisely obtained by an expansion 
around the zero value of the supersymmetry-breaking parameter.
In this case, 
as long as the $R_{12} \to 0$ (de-)compactification limit can be taken, 
it is possible to go to a supersymmetric limit, 
otherwise, supersymmetry would be broken.

\subsection*{\sl a possible scenario}

The fact that under
compactification/decompactification it is possible to
go to a supersymmetric limit in eleven dimensions
suggests that a  naive picture of the situation could be
the following: the theory lives on a twelve-dimensional ball,
whose boundary is an eleven dimensional sphere: this is fine,
since we know that anyway the theory must have a horizon.
Supersymmetry is then trivially broken owing to the compactness of this
space. Whether the space is really  twelve dimensional, or
we have rather to deal with just an eleven dimensional curved space,
it does not make at this stage a big difference: for all
what we know, it could well be that the twelfth coordinate is just
a curvature parameter. From a mathematical point of view,
it is always possible to embed a curved hypersurface into
a (fictitious) higher dimensional flat space.
Although not necessarily twelve dimensional from the very beginning, the 
theory would then become anyway ``effectively'' twelve dimensional
as soon as its space gets sufficiently ``curved'' during the
process of ``singularization'' produced by entropy evolution.
The supersymmetric limit is just the one in which we send the radius
(related therefore to the curvature) to zero/infinity.
There, we recover an eleven dimensional, locally flat surface, on which
there is (at least locally), supersymmetry.  
When however this hyperspace is sufficiently curved
(${\cal N}_4=1$ situations), consistency
of the theory implies that also the ``radial'' coordinate is twisted.
This forbids taking a true decompactification limit, and
supersymmetry  is necessarily broken. This phenomenon does not
appear ``locally'' in the hypersphere: from the point of view
of the hypersphere supersymmetry is perturbatively preserved,
because perturbation is defined as an expansion around the zero
value of the ``radial'' coordinate. The breaking of supersymmetry is
therefore non-perturbative at that stage. However, one can also chose
to make a different identification of the string (or M-theory) 
coordinates inside the twelve-dimensional ball, thereby having the
possibility of constructing explicitly non-supersymmetric theories.

\subsection*{\sl the origin of confinement}

Under decompactification of the radial coordinate, 
half of the theory is ``washed out''.
Namely, we obtain a limit in which the theory can be described entirely
in terms of weakly coupled states: the part which is 
T-dual under inversion of the
radial coordinate, that would therefore be strongly coupled with
respect to the one we keep, is decoupled. When instead
the ``radial'' coordinate is twisted, we cannot take a limit
in which we decouple one of the two mutually strongly coupled parts.
Therefore, the theory cannot be entirely described in terms of weakly  
coupled states: in this case there are 
sectors which are ``S-dual'' with respect to each other:
we have at the same time the presence of weakly and strongly coupled
sectors. From this point of view, we understand now why
in ${\cal N}_4=1$ vacua supersymmetry
is always non-perturbatively broken due to gaugino condensation
in a strongly coupled sector, as suggested in Ref.~\cite{striality}:
in these vacua this coordinate is necessarily twisted.
This is also  the reason why in our World there is quark confinement.

\subsection*{\sl a note on space-time geometry}

We have seen  that the existence of a non-vanishing cosmological 
constant is a truly quantum effect, due to the finiteness of our horizon.
Asymptotically, namely, in a space in which the horizon is
at infinity, the cosmological constant vanishes, and the space-time becomes 
flat. In this framework, we find inappropriate the search of supergravity
configurations, or more in general for classical geometric configurations,
reproducing the characteristics of a non-compact space with positive
cosmological constant. This does not correspond in fact
to the actual physical situation, which better corresponds instead
to a ``quantum geometry''. Moreover, as we have seen, the string 
coupling, i.e. the coupling of the quantum theory of our world, is 
of order 1. We are therefore out of the supergravity domain (supergravity,
even with broken supersymmetry, is just a first approximation, 
valid in certain cases at the weak coupling), and we are not allowed
to expect that the real physical situation should find a description
in that framework.

\section{\bf Conclusions}
\label{conclusions}

In this work, we presented a proposal for a set up
in which String Theory is implemented by a dynamical principle, 
a ``stringy'' version of the second law of thermodynamics, that
provides us with a ``rationale'' through the plethora of possible
string configurations. This allows us to ``solve'' the theory
during its evolution,
and identify the configuration corresponding to the world as we observed
it as the only consistent solution at the present time. 
Although a mathematical framework in which to properly set 
the problem and make rigorous computations has still to be
worked out, nevertheless, by using different types
of approximations, we derived phenomenological predictions, 
both qualitative and up to a certain extent also quantitative. 
Among these, the expected order of magnitude of neutrino masses.
Among the most striking ones, the non-existence of the Higgs particle,
in fact not needed in order to generate masses in this set up,
and the non-existence of supersymmetric partners of the observed
particles and fields. More precisely, ``Higgs'' and supersymmetric
modes are to be found at the Planck scale, a scale at which
we are perhaps not anymore allowed to speak in terms of
``particles'' and ``fields''. Precisely a mass gap of the
order of the Planck scale between the observed particles and their
superpartners is what produces the observed value of the 
cosmological constant. The latter turns out to be nothing but
a quantum effect, ``measuring'' our horizon of observations.
It therefore decreases as space-time expands. Perhaps
the most unexpected result is the fact that the Universe
knows only about one fundamental scale,
the Planck scale; all mass scales
below the Planck mass depend on the point in the time evolution
of the Universe: light masses precisely arise as a consequence of the existence
of a horizon to our observation;
they are related to the ``time'' coordinate,
and decrease as the observable horizon increases. 
This effect, although extremely slow at the present day,
is by itself able to explain the accelerated
expansion of the Universe, and its direct observation (or non-observation)
would be a key test of our proposal, both in order to support it
or to rule it definitely out.

Inspection of various perturbative and non-perturbative
properties of string vacua lead then us to
make some guess about the theory underlying 
string theory. In our framework, the breaking of supersymmetry
and quarks confinement receive a natural explanation,
as the effect of a coordinate ``twisting'' 
necessarily implied by the second law of thermodynamics.

This work does not have to be considered as conclusive,  but rather
on the contrary, as a first step toward a ``definition'' of the theory.
To this purpose, absolute priority has to be given to the
setting of the correct mathematical tools enabling a rigorous
and quantitative investigation of what we called ``the most singular
configuration'' of the string space. If we decide that
the history of our Universe starts at the Planck time/scale,
this can be regarded as the true ``starting point'', or initial condition. 
After all, our interpretation of the string path through 
the various possible vacua to reach this configuration can be questionable, 
and the arguments be refined. 
It remains however that all the subsequent history seems to originate from
there.

\vskip 1.5cm
\centerline{\bf Bibliography}

\noindent
The following list of references is largely incomplete.
Correctly quoting all the works related to what I have 
presented would be a hard task. I therefore decided
to list only those references which are immediately
needed in order to complete the discussion. I apologize for the 
many omissions.

\vspace{1.5cm}



\providecommand{\href}[2]{#2}\begingroup\raggedright\endgroup

\end{document}